\documentclass[conference]{IEEEtran}
\IEEEoverridecommandlockouts
\usepackage{amsmath,amssymb,amsfonts}
\usepackage{algorithm}
\usepackage{algpseudocode} %
\usepackage{graphicx}
\usepackage{textcomp}
\usepackage{xcolor}
\def\BibTeX{{\rm B\kern-.05em{\sc i\kern-.025em b}\kern-.08em
    T\kern-.1667em\lower.7ex\hbox{E}\kern-.125emX}}

\usepackage{nicefrac}
\usepackage{fancyref}
\usepackage{bm}		%
\usepackage{caption}
\usepackage{hyperref}

\usepackage[normalem]{ulem} %
\usepackage{multirow}

\usepackage{mathtools}
\mathtoolsset{showonlyrefs}

\usepackage{siunitx}
\makeatletter
\@ifpackagelater{siunitx}{2021/05/17} %
{
	\sisetup{per-mode=fraction,
		inter-unit-product=\ensuremath{{\cdot}},
		exponent-product=\ensuremath{\cdot},
		fraction-function=\nicefrac}
}
{	
	\sisetup{per=frac,
		fraction=nice,
		decimalsymbol=comma,
		binary-units=true,
		loctolang={DE:ngerman,UK:english}}
}%
\makeatother

\newcommand{\ttt}{\! \left( t \right)}
\newcommand{\tf}{ $t_\text{f}$ \ }
\newcommand{\tft}{\! \left( t_\text{f} \right)}
\newcommand{\tftt}{\! \left( t_\text{f,t} \right)}
\newcommand{\tzt}{\! \left( 0 \right)}

\newcommand{\xt}{\bm{x} \ttt}
\newcommand{\ut}{\bm{u} \ttt}

\newcommand{\xtutt}{\! \left( \xt \! , \, \ut \! , \, t\right)}

\newcommand{\tct}{\! \left( \cdot \right)}
\newcommand{\nel}{n_{\text{el}}}
\newcommand{\jmax}{j_{\text{max}}}
\newcommand{\amax}{a_{\text{max}}}

\renewcommand{\tf}{t_{\text{f}}}
\newcommand{\phiabs}{\varphi_{\text{abs}}}
\newcommand{\phif}{\varphi_{\text{f}}}
\newcommand{\vellm}[1]{\bm{\ell}_{\text{m#1}}}

\newcommand{\wrt}{w.r.t. }%

\newcommand{\ocpJ}{\mathtt{OCP}\text{-}\mathtt{{J}}}
\newcommand{\ZV}{\mathtt{ZV}}

\newcommand{\SCurve}{\mathtt{S}\text{-}\mathtt{{Curve}}}

\newcommand{\Fir}{\mathtt{FIR}}

\newtheorem{theorem}{Theorem}
\newtheorem{conjecture}{Conjecture}

\begin{document}

\title{Calculation of time-optimal motion primitives for systems exhibiting oscillatory internal dynamics%
\thanks{This work has been submitted to the IEEE for possible publication. Copyright may be transferred without notice, after which this version may no longer be accessible.}
}

\author{\IEEEauthorblockN{1\textsuperscript{st} Thomas Auer}
	\IEEEauthorblockA{\textit{IACE - UMIT TIROL}\\
		Hall in Tirol, Austria \\
		thomas.auer@umit-tirol.at}
	\and
	\IEEEauthorblockN{2\textsuperscript{nd} Frank Woittennek}
	\IEEEauthorblockA{\textit{IACE - UMIT TIROL}\\
		Hall in Tirol, Austria \\
		frank.woittennek@umit-tirol.at}
}

\maketitle

\begin{abstract}
An algorithm for planning near time-optimal trajectories for systems with an oscillatory internal dynamics has been developed in previous work. It is based on assembling a complete trajectory from motion primitives called jerk segments, which are the time-optimal solution to an optimization problem. To achieve the shortest overall transition time, it is advantageous to recompute these segments for different acceleration levels within the motion planning procedure. This publication presents a numerical calculation method enabling fast and reliable calculation. This is achieved by explicitly evaluating the optimality conditions that arise for the problem, and further by reducing the evaluation of these conditions to a line-search problem on a bounded interval. This reduction guarantees, that a valid solution if found after a fixed number of computational steps, making the calculation time constant and predictable. Furthermore, the algorithm does not rely on optimisation algorithms, which allowed its implementation on a laboratory system for measurements with the purpose of validating the approach. \\
  
\end{abstract}

\def\abstractname{Note to Practitioners}
\begin{abstract}
	This publication focuses on trajectory planning for systems where oscillation at the end of the motion must be avoided. An example of such systems are high-precision pick-and-place systems in the semiconductor industry. The focus of this publication is on calculating motion profiles complete trajectories can be easily assembled from. These primitives are each planned to provide an oscillation-free transition between different acceleration levels. A supplementary publication shows the calculations required for a one-dimensional pick-and-place system. The main advantage of the calculation presented here lies in its ease of implementation, where a time-optimal solution for the motion profiles is calculated without the need for sophisticated optimization algorithms, allowing straightforward implementation on a PLC. The formulation also ensures that the solution is computed within a known time. \\
\end{abstract}

\begin{IEEEkeywords}
	Motion-planning, Optimization methods, Motion Control
\end{IEEEkeywords}

\section{Introduction}
Pick-and-place processes in the electronics industry require high accuracy and speed. Current developments in further integration and miniaturisation require high accuracy in component placement \cite{LindaS_Wilson2023,Lau2025,Ikegami2024,Haneda2024}. High accuracy is required for proper component placement to ensure electrical connections work as intended. To reduce cost, short transitions are desirable. Motion forces on the machine frame combined with finite stiffness lead to oscillation and inaccuracy in component placement. The oscillation of the machine frame can be greater than the accuracy requirements, which are fractions of $\si{\micro\meter}$'s \cite{Lau2025,Ikegami2024,Haneda2024}. Even if accuracy requirements do not dictate the removal of oscillation, high oscillation can lead to excessive wear in the machine, so it should be avoided \cite{Esau2022,Bilal2023}. This can be achieved by specially planned motion profiles, also called trajectories \cite{Auer2023IFAC,Auer2024Case}.

\subsection{Related work}
In general, trajectory shapers can be used very effectively to remove oscillations from systems \cite{Kruk2023}. A commonly used trajectory shaper (often called a trajectory filter) is the $\ZV$-filter \cite{Singhose1990,Singhose1994}, which relies on knowledge of the system dynamics and parameters. Parameter uncertainty due to operational loads \cite{Zhou2015} or model uncertainty from linear approximation \cite{Maghsoudi2017} lead to residual oscillations at the end of the transition \cite{Auer2023IFAC}. Shapers specifically optimised for parameter insensitivity exist and are well studied \cite{Pao1998,Kasprowiak2022,Kang2019,Singhose1997_EI_shaper,Singer1999,Singhose1994,Vaughan2008}. Insensitivity to parameter changes is usually associated with longer shaper durations and thus slower movements. So-called negative impulse shapers can be used to shorten the transition times. Special care must be taken to properly account for the kinematic constraints of the actuators \cite{SinghoseMarch1996, Sorensen2008}. It has been found by \cite{Cao2021} that negative impulse shapers can lead to a reduction in transition time, while providing less sensitivity to $\ZV$-filters for selected regions of parameter uncertainty. Other approaches focus on adjusting the jerk to avoid oscillations altogether without using a filter \cite{Meckl1998, Kim2018}. It has been shown that damping can be considered \cite{Bearee2014,Biagiotti2015} and that it can be extended to the full trajectory \cite{Biagiotti2019,Biagiotti2020,Biagiotti2021,Yalamanchili2024}. This reduces overall transition time at the expense of sensitivity to parameter uncertainty \cite{tau_ocpJ_assembly_part1}. Unlike algorithms such as \cite{Haschke2008,Pham2014,Beul2016,Berscheid2021,Wang2024}, which can be used to plan trajectories, the approach presented in this paper considers not only the state variables of the actuated joints, but also the kinematic state variables of the free oscillation of the system.

\bigskip
A method for assembling trajectories from pre-planned motion primitives (called jerk segments) is presented in \cite{tau_ocpJ_assembly_part1}, where the potential for improvement in terms of transition time is highlighted. Compared to $\ZV$-filters  \cite{Singhose1990,Singhose1994}, a reduction in transition time can be achieved. Compared to $\Fir$-based methods \cite{Biagiotti2015,Biagiotti2019,Biagiotti2020,Biagiotti2021,Yalamanchili2024} the approach offers advantages in terms of parameter sensitivity. The current publication focuses on a direct method to compute the jerk profiles themselves (solving the problem mentioned by \cite{Dijkstra2007}, that arises from optimization-based approaches).

\subsection{Contribution of this paper}
The main contribution of this paper is the derivation of an explicit solution of the optimal control problem defining the jerk segments. The presented method requires only elementary functions ($\sin\tct$, $\cos\tct$, etc.) and does not rely on optimization libraries, allowing implementation on a PLC. The formulation also guarantees that a solution is found after a fixed number of computational steps. The proposed method is validated experimentally with measurements (both the jerk segments and the trajectories were computed directly on the PLC itself). The presented method can be applied to systems with weakly damped oscillatory internal dynamics, described by an oscillation frequency $\omega_{\text{d}}$ and damping $\delta$.

\bigskip
\subsection{Organisation of paper}
This paper is organized into the following sections. The system model is introduced in \autoref{sec:model_and_paramters}. The method of assembling the trajectories from jerk segments is sketched briefly in \autoref{sec:calc_full_trajectory}. The efficient method of calculating the individual jerk segments for the full trajectories is presented in \autoref{sec:efficient_algorithm}. Subsequently in \autoref{sec:comp_to_others} the proposed approach is compared to other methods conceivable to solve the stated problem. Measurement results for validation and a description of the laboratory system are given in \autoref{sec:comp_to_others}. The results of the publication are discussed in \autoref{sec:discussion}. A summary and outlook for relevant follow-up research is given in \autoref{sec:summary_and_comparison}.

\section{Model and mathematical background}\label{sec:model_and_paramters}

\begin{figure}[!ht]
	\centering
	\def\svgwidth{0.95\linewidth}
	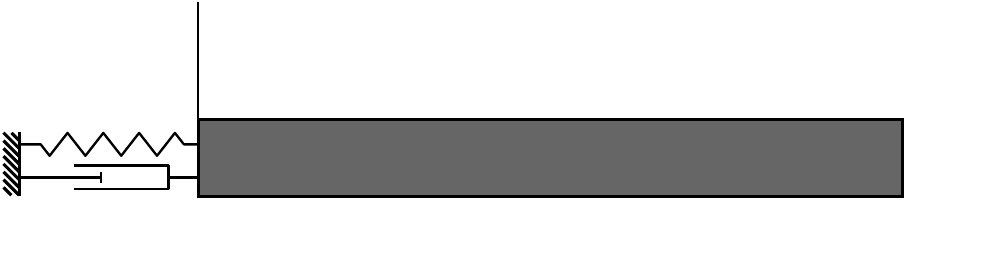
	\caption{Symbolic representation of one axis of the pick-and-place machine.}
	\label{fig:model_of_the_system}
\end{figure}

As mentioned in \cite{tau_ocpJ_assembly_part1}, the pick-and-place machine is modelled as a lumped system with two masses (mass of the baseframe $m_\text{b}$ and mass of the slider $m_\text{s}$), a spring with stiffness $k$ and a viscous damping element with damping parameter $d$. The schematic representation of the system is shown in \autoref{fig:model_of_the_system}.
This was chosen over a continuum-mechanics description, which it was deemed sufficient \cite{DissKroni2022, Auer2024Case}, because the essential oscillation of the machine frame can be captured by the first mode of the flexible system. Momentum balance allows to derive the equations of motion%
\begin{subequations}\label{eq:dynamical_equations_of_system}
	\begin{align}
		m_\text{b}  \ddot{x}\ttt + d \, \dot{x}\ttt + k \, x\ttt &= -F\ttt \, \text{,} \label{eq:movement_equation_for_base} \\
		m_\text{s} \! \left(\ddot{x}\ttt + \ddot{z}\ttt\right) &= F\ttt \, \text{.} \label{eq:force_acting_in_system}
	\end{align}
\end{subequations}
and the parameters used in the numerical studies are listed in \autoref{tab:table_listing_machine_parameters}.
\begin{table}[!ht]
	\captionsetup{width=\linewidth}
	\caption{Parameters and resulting eigenfrequencies}
	\vspace{-1em}
	\renewcommand{\arraystretch}{1.25}
	\centering
	\begin{tabular}{|l|l|l|l|}
		\hline
		$m_\text{s} = \SI{25}{\kilo\gram}$ & $m_\text{b} = \SI{500}{\kilo\gram}$ &
		$k = \SI{15e6}{\newton\per\metre}$ & $d = \SI{5e3}{\kilo\gram\per\second}$ \\ \hline
		\multicolumn{2}{|l|}{$f_\text{0} = \SI{26.9}{\hertz}$} & \multicolumn{2}{l|}{$f_\text{d} = \SI{26.8914}{\hertz}$} \\ \hline
	\end{tabular}
	\label{tab:table_listing_machine_parameters}
\end{table}
The kinematic constraints listed in \autoref{tab:table_listing_kinematic_contraints} correspond to an exemplary production machine.
\begin{table}[!ht]
	\captionsetup{width=\linewidth}
	\caption{Kinematic constraints of slider.}
	\vspace{-1em}
	\renewcommand{\arraystretch}{1.3}
	\centering
	\begin{tabular}{|l|l|l|}
		\hline
		$\left|\dot{z}\ttt\right| \le v_\text{lim}$ & $\left|\ddot{z}\ttt\right| \le a_\text{lim}$ &
		$\big|z^{\left(3\right)}\ttt\big| \le j_\text{lim}$ \\ \hline
		$v_\text{lim} = \SI{1.5}{\meter\per\second}$ & $a_\text{lim} = \SI{20}{\meter\per\square\second}$ &
		$j_\text{lim} = \SI{800}{\meter\per\second\cubed}$ \\ \hline
	\end{tabular}
	\label{tab:table_listing_kinematic_contraints}
\end{table}
It is advantageous to use a dynamic extension \cite{tau_ocpJ_assembly_part1} with $z^{\left(3\right)}\ttt$ as input for the calculations required later on. Adding \eqref{eq:movement_equation_for_base} and \eqref{eq:force_acting_in_system} together to remove $F\ttt$, using ${m_\text{g} = m_\text{s} + m_\text{b}}$ and selecting ${u\ttt = z^{\left(3\right)}\ttt}$ as the input of the dynamic extension leads to the state-space representation
\begin{align}\label{eq:opt:sys}
	\bm{\dot{x}} &= %
	\underbrace{\begin{bmatrix}
		0 & 1 & 0& 0& 0 \\
		-k^\star & -d^\star &0&0&-m^\star \\
		0&0&0&1&0 \\
		0&0&0&0&1 \\
		0&0&0&0&0
	\end{bmatrix}}_{\bm{A}} \cdot \begin{bmatrix} x \\ \dot{x} \\ z \\ \dot{z}  \\ \ddot{z} \end{bmatrix} + 
	\underbrace{\begin{bmatrix} 0 \\ 0 \\ 0 \\ 0  \\ 1 \end{bmatrix}}_{\bm{b}} \! z^{\left(3\right)} \\
	k^\star &= \nicefrac{k}{m_\text{g}} \qquad d^\star = \nicefrac{d}{m_\text{g}} \qquad m^\star = \nicefrac{m_\text{s}}{m_\text{g}}
\end{align}
with the state and input
\begin{align}
	\bm{x} &= \left[x, \dot{x}, z, \dot{z}, \ddot{z}\right]^T \, \text{,} & u &= z^{\left(3\right)} \, \text{.} \label{eq:sys_state_for_Axbu}
\end{align}

\section{Method explained: $\ocpJ$}\label{sec:calc_full_trajectory}
\begin{figure}[!ht]
	\centering
	\def\svgwidth{0.95\linewidth}
	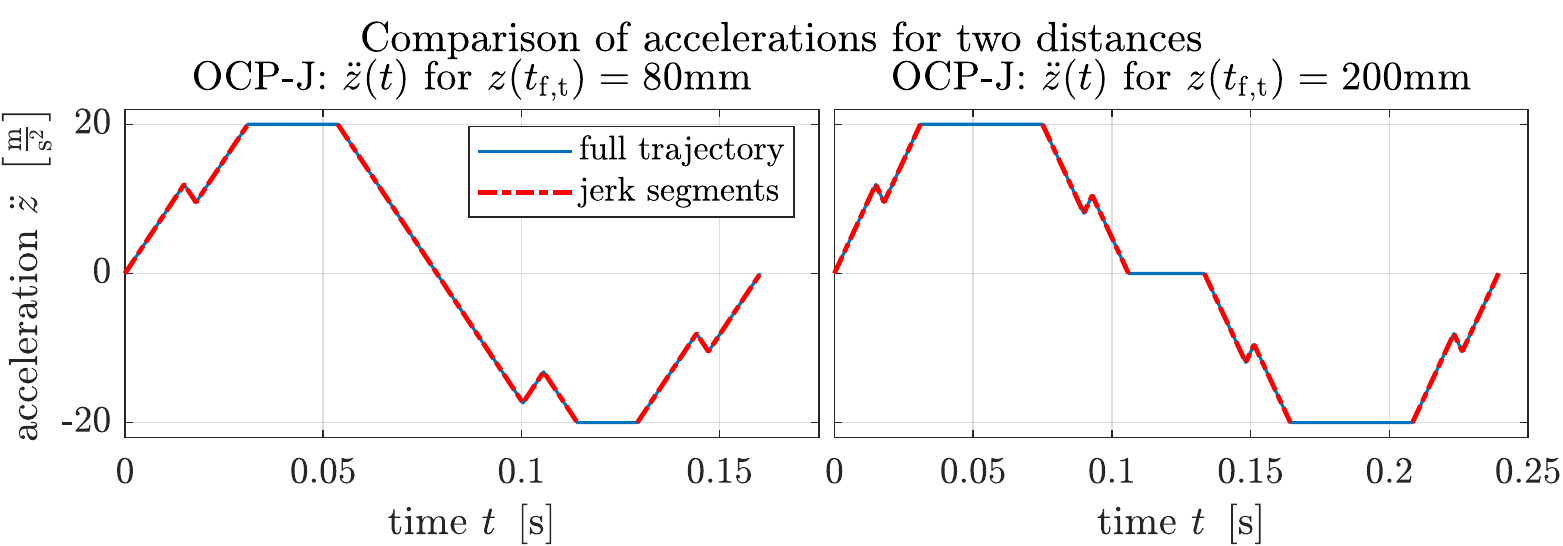 %
	\caption{Two $\ocpJ$ trajectories demonstrating how a full trajectory is assembled from the jerk segments (picture taken from \cite{tau_ocpJ_assembly_part1}).}%
	\label{fig:two_OcpJ_trajectories_for_demo_of_method}
\end{figure}

\begin{figure}[!ht]
	\centering
	\def\svgwidth{0.95\linewidth}
	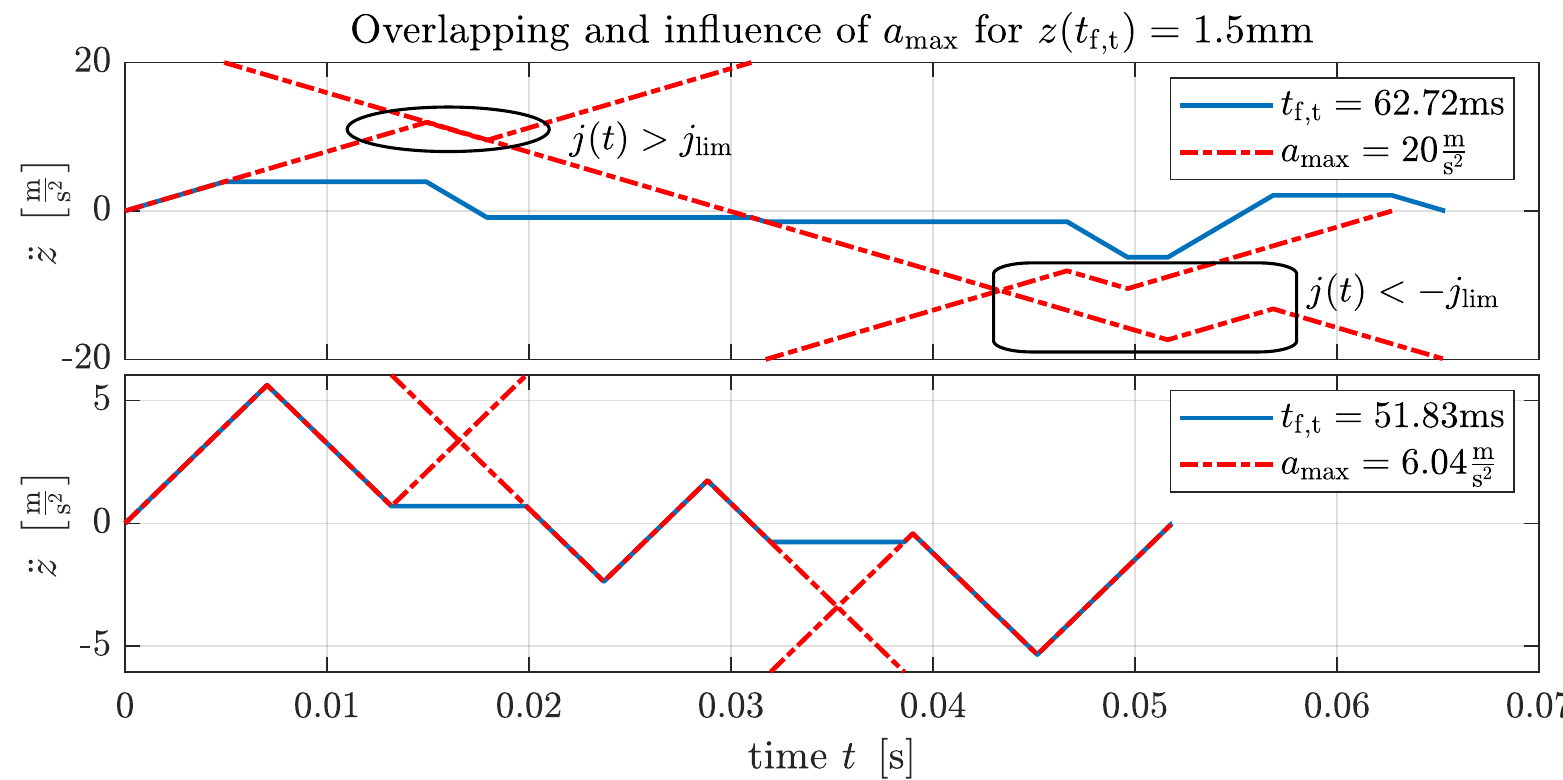 %
	\caption{Showing the overlapping without adjustment of $a_\text{max}$ for a transition distance of $z\tftt=\SI{1.5}{\milli\meter}$ to show the advantage of recalculation of jerk segments in order to reduce the transition time. Picture taken from \cite{tau_ocpJ_assembly_part1}.}
	\label{fig:showing_overlapping_1p5mm}
\end{figure}

This section gives an overview of $\ocpJ$ and shows how the content presented in this publication is used in the $\ocpJ$ method. Since this is the main content of \cite{tau_ocpJ_assembly_part1}, this section is kept short and serves only as an introduction to show how the results of that publication can be used. %
The essential idea is, that a complete trajectory is assembled from individual motion primitives (called jerk segments), as shown in \autoref{fig:two_OcpJ_trajectories_for_demo_of_method}.
In order to ensure transitions to the end point in minimum time, these jerk segments must themselves be time-optimal transitions, as shown in \cite{tau_ocpJ_assembly_part1}. The advantage an efficient computation offers is highlighted in \autoref{fig:showing_overlapping_1p5mm}.
The top trajectory has been calculated with $a_\text{max} = \SI{20}{\meter\per\second\squared}$ and the bottom trajectory with $a_\text{max} = \SI{6.04}{\meter\per\second\squared}$. A full trajectory is a concatenation of several jerk segments with intermediate time intervals of zero jerk (see also \autoref{fig:two_OcpJ_trajectories_for_demo_of_method}). For very short transition distances, the lengths of these intervals can be negative, resulting in overlapping jerk segments, as visualized in \autoref{fig:showing_overlapping_1p5mm}. Such overlaps can lead to violations of the jerk constraints (see the upper plot in \autoref{fig:showing_overlapping_1p5mm}). If the overlap is as large as in the top plot, this can be problematic, as it leads to the jerk constraint being violated in the regions marked in the plot. Reducing the maximum acceleration used to plan the jerk segments (cf.~\cite{tau_ocpJ_assembly_part1} for details) serves two purposes. First, violating the kinematic constraints can be avoided by reducing the overlap of the jerk segments. Second, the overall transition time can be reduced as shown in \autoref{fig:showing_overlapping_1p5mm}. This is analysed in more detail in \cite{tau_ocpJ_assembly_part1}. To summarize, recalculating the jerk segments can be beneficial in reducing the overall transition time and resolving kinematic constraint violations. As stated in the introduction, the algorithm/methodology for efficiently and reliably calculating these jerk segments is the main contribution of this thesis and is presented in \autoref{sec:efficient_algorithm}.

\section{Efficient algorithm to calculate the jerk segments}\label{sec:efficient_algorithm}
As demonstrated in \cite{tau_ocpJ_assembly_part1} and shown in \autoref{fig:showing_overlapping_1p5mm}, recalculation of the jerk segments is required to ensure minimal transition times $t_\text{f,t}$. %
This section introduces and presents the algorithm that ensures fast and reliable calculation of those jerk segments. They can be obtained as the solution of an optimization problem with cost
\begin{subequations}\label{eq:optim_ctrl_problem}
	\begin{align}
		\min_{\bm{x}\in\mathbb{R}^n, u\in\left[-j_\text{max},j_\text{max}\right]} & J\xtutt \, \text{,} \label{eq:minJ_opt_problem} \\
		J\xtutt &= \int\limits_{0}^{t_\text{f}}1\text{dt} = t_\text{f} \, \text{.} \label{eq:J_cost_opt_problem}
	\end{align}
\end{subequations}
At the end of the segment $t_\text{f}$ ($t_\text{f}$ always denotes the terminal time for a jerk segment and $t_\text{f,t}$ the terminal time for the entire trajectory composed of multiple jerk segments), a certain acceleration (called $a_\text{max}$) should be reached, while the oscillation of the base $x\ttt$ must be equal to zero. The acceleration $\ddot{z}\ttt$ of the slider is constant at this point, resulting in a constant displacement of the base $x\ttt$. Since the model and parameters are known, this results in a transition of the system state variables \eqref{eq:sys_state_for_Axbu} from
\begin{equation}\label{eq:startState_system_ocpJ_jerkSegment}
	\bm{x}\tzt = \left[x, \dot{x}, z, \dot{z}, \ddot{z}\right]^T = \left[0,0,0,0,0\right]^T
\end{equation}
to the partly constraint end state (rest of system state variables are left open)
\begin{equation}\label{eq:endState_system_ocpJ_jerkSegment}
	\bm{x}\tft = \left[x, \dot{x}, \ddot{z}\right]^T = \left[-\frac{a_\text{max} m_\text{s}}{k},0,a_\text{max}\right]^T \, \text{.}
\end{equation}
Twice continuously differentiable trajectories $t\mapsto z\ttt$, which are piecewise constant in $z^{\left(3\right)}\ttt$ are considered, since they allow to limit the velocity, acceleration and the jerk. 
Those equations allow to use Pontryagin's Maximum Principle (PMP) \cite{pontriagin1964mathematische} to solve the problem time-optimally and show the necessary optimality conditions.

\subsection{Optimality conditions}\label{SubSec:PMP_for_jerk_segments}
\begin{figure}[!ht]
	\centering
	\def\svgwidth{0.95\linewidth}
	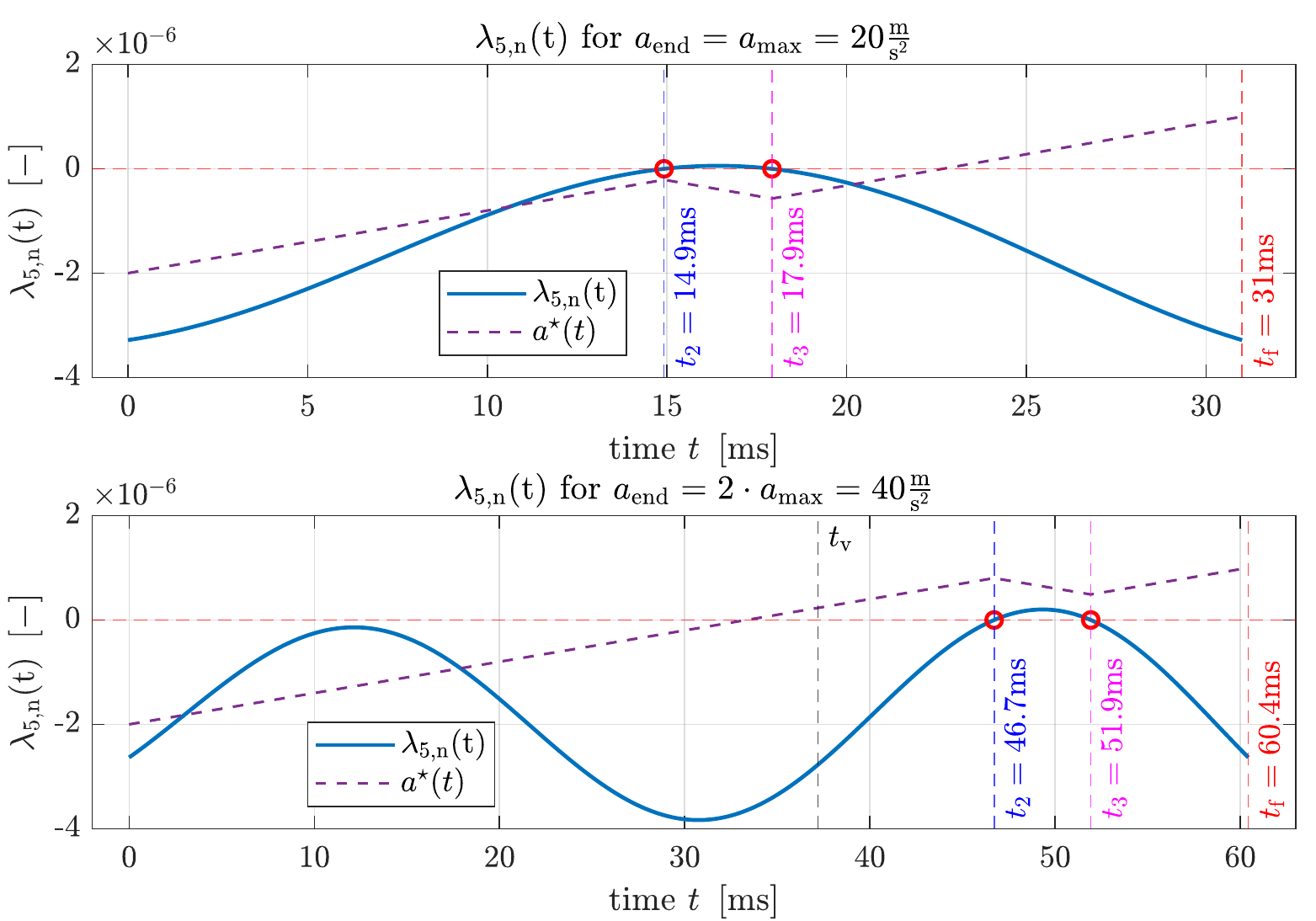 %
	\caption{Results for $\lambda_\text{5,n}\ttt$ for the two jerk profiles required in the trajectories shown in \autoref{fig:two_OcpJ_trajectories_for_demo_of_method}. Included are the acceleration profiles $\ddot{z}\ttt$ of the jerk segments for comparison.}
	\label{fig:exemplary_solution_lambda1_and_5}
\end{figure}

In the following, Pontryagin's maximum principle \cite{pontriagin1964mathematische} is employed in order to solve the optimization problem \eqref{eq:optim_ctrl_problem} \noeqref{eq:minJ_opt_problem,eq:J_cost_opt_problem}.%
With the co-state (also called adjoint state) $\bm{\lambda}\ttt=(\lambda_1\ttt,\dots,\lambda_5\ttt)^\intercal$ the Hamiltonian \cite{pontriagin1964mathematische,Athans2007,Hocking1991} required to formulate the optimality conditions is given by
\begin{equation}
	H\!\left(\bm{\lambda},\bm{x},u\right) = 1+\bm{\lambda}^\intercal\left(\mathbf{A}\bm{x}+\mathbf{b} u\right)=u\lambda_5+\mathcal{R}\left(\bm{x}, \bm{\lambda}\right)
 \end{equation}
 with
\begin{equation}
	\mathcal{R}\left(\bm{x}, \bm{\lambda}\right)=1+\lambda_1 \, \dot{x} + \lambda_2 \! \left(-k^\star \! x -d^\star \! \dot{x} -m^\star \! \ddot{z}\right) + \lambda_3 \, \dot{z} + \lambda_4 \, \ddot{z} \, \text{.}
 \end{equation}
According to the maximum principle $u\ttt$ must satisfy\footnote{For convenience we do not explicitly distinguish the time optimal trajectory.}
\begin{equation}
	\begin{split}
		u\ttt&=\arg\min_{\bar u\in[-j_\text{max},j_\text{max}]} H(\bm{x}(t),\bm{\lambda}(t),\bar u)\\
		&=\arg\min_{\bar u\in[-j_\text{max},j_\text{max}]} \lambda_5(t) \, \bar u + \mathcal{R}\!\left(\bm{x}(t), \bm{\lambda}(t)\right) \, \text{.}
	\end{split}
\end{equation}
Hence,
\begin{equation}\label{eq:PMP_input_required_from_Hamiltonian}
	u\ttt = 
	\begin{cases}
		-j_\text{max}, & \text{if} \quad \lambda_{5}\ttt > 0 \, \text{,} \\
		\hphantom{-}j_\text{max}, & \text{if} \quad \lambda_{5}\ttt < 0 \, \text{.}
	\end{cases}
\end{equation}
Therefore, the time-optimal control law corresponds to a bang-bang behaviour. 
Further (necessary) optimality conditions are constituted by the adjoint equations
\begin{equation}
	{\dot{\lambda}}_i = -\frac{\partial H}{\partial x_i}, \quad \ i=1,..., 5,
\end{equation}
i.e.,
\begin{subequations}
	\begin{align}
		\dot{\lambda_1} &= \lambda_2 \, k^\star \, \text{,} \label{eq:required_for_calc_of_lambdaU_1} \\
		\dot{\lambda_2} &= \lambda_2 \, d^\star - \lambda_1 \, \text{,} \label{eq:required_for_calc_of_lambdaU_2} \\
		\dot{\lambda_3} &= 0 \, \text{,} \label{eq:required_for_calc_of_lambdaU_3} \\
		\dot{\lambda_4} &= -\lambda_3 \, \text{,} \label{eq:required_for_calc_of_lambdaU_4} \\
		\dot{\lambda_5} &= \lambda_2 \, m^\star - \lambda_4 \, \text{.} \label{eq:required_for_calc_of_lambdaU_5}
	\end{align}
\end{subequations}
Finally, initial and terminal conditions for the state variables in $\bm{x}\ttt$ as well as the transversality conditions associated with the unconstrained final values of the third and fourth system state variable $\left(z\tft, \dot{z}\tft\right)$ read\footnote{Note that the optimal control problem can be equivalently posed on $\mathbb{R}^3$ with state $(x(t),\dot x(t),\ddot{z}(t))^\intercal$. However,
for the sake of readability the full state is retained here, which does not increase complexity.} %
\begin{subequations}\label{eq:boundary_constraints_PMP_for_system}
	\begin{align}
		x\tzt &= 0 \, \text{,} & x\tft &= -\frac{a_\text{max} m_\text{s}}{k} \, \text{,} \label{eq:BC_xtzt_xtft_segment} \\
		\dot{x}\tzt &= 0 \, \text{,} & \dot{x}\tft &= 0 \, \text{,} \label{eq:BC_xptzt_xptft_segment} \\
		z\tzt &= 0 \, \text{,} & \lambda_3\tft &= 0 \, \text{,} \label{eq:BC_ztzt_l3tft_segment} \\
		\dot{z}\tzt &= 0 \, \text{,} & \lambda_4\tft &= 0 \, \text{,} \label{eq:BC_zptzt_l4tft_segment} \\
		\ddot{z}\tzt &= 0 \, \text{,} & \ddot{z}\tft &= a_\text{max} \, \text{.} \label{eq:boundary_constraints_PMP_for_system_zpp}
	\end{align}
\end{subequations}
Therein, the constraints on the final state result from $\dot{x}\tft=\ddot{x}\tft=0$, which correspond to an equilibrium of the oscillatory internal dynamics, while taking into account the terminal constraint on $\ddot{z}$, i.e., $\ddot{z}\tft=a_{\text{max}}$. The adjoint equations \eqref{eq:required_for_calc_of_lambdaU_3} and \eqref{eq:required_for_calc_of_lambdaU_4} in connection the transversality conditions \eqref{eq:BC_ztzt_l3tft_segment} and \eqref{eq:BC_zptzt_l4tft_segment} result in 
\begin{equation}\label{eq:lambda_3_lambda_4_zero}
	\lambda_3 \ttt = \lambda_4 \ttt = 0 \, \text{.}
\end{equation}
In order to calculate the required input in accordance with \eqref{eq:PMP_input_required_from_Hamiltonian}, an equivalent second order differential equation for $\lambda_{2}$ is deduced from \eqref{eq:required_for_calc_of_lambdaU_1} and \eqref{eq:required_for_calc_of_lambdaU_2}:
\begin{equation}
	\ddot{\lambda}_2 \ttt = d^\star \dot{\lambda}_2 \ttt - k^\star \lambda_2 \ttt \, \text{.}
\end{equation}
Using $\delta = \nicefrac{d^\star}{2}$, $\omega_0^2 = k^\star$ and $\omega_\text{d}^2 = k^\star-\delta^2$ the general solution of this equation is
\begin{equation}\label{eq:lambda_2_equation_PMP}
	\lambda_2 \ttt = A_2 \, e^{\delta \, t} \, \cos \! \left(B_2 + \omega_\text{d} \, t\right) \, \text{.}
\end{equation}
Moreover, in view of \eqref{eq:lambda_3_lambda_4_zero} and \eqref{eq:required_for_calc_of_lambdaU_5},
$\lambda_5$ is obtained by integrating \eqref{eq:lambda_2_equation_PMP}:
\begin{equation}
	\lambda_5\ttt = A_2 \frac{m_\text{s}}{k} e^{\delta t } \left[\delta \cos\! \left(B_2 + \omega_\text{d} \, t\right) + \omega_\text{d}\sin\! \left(B_2 + \omega_\text{d} \, t\right)\right] + C_2 \, \text{.}
\end{equation}
This expression is rewritten with appropriate constants:
\begin{equation}
	\lambda_5\ttt = A_1 e^{\delta t} \sin\! \left(B_1 + \omega_\text{d} \, t\right) + C_2 \, \text{.}
\end{equation}
Since the problem exhibits free end time, taking the end conditions \eqref{eq:boundary_constraints_PMP_for_system} into account, an additional transversality condition is given by
\begin{subequations}
	\begin{align}
		0&= H\!\left(\bm{x}\tft, \bm{\lambda}\tft, u\tft\right)\\
		&=\lambda_{5}\tft j_\text{max} +1 - \lambda_2 \tft m^\star a_\text{max} \, \text{.}
	\end{align}
\end{subequations}
In the generic case,
\begin{equation}
	\left(\delta-k^\star a_{\text{max}}\right) \cos\! \left(B_2 + \omega_\text{d} \, \tf\right) + \omega_\text{d}\sin\! \left(B_2 + \omega_\text{d} \, \tf\right)\ne 0 \, \text{,}
\end{equation}
the transversality condition can always be met by appropriately choosing $A_2$, which in turn determines $A_1$. Otherwise,
\begin{equation}
	C_2 = -\frac{1}{\jmax}.
\end{equation}
In the following, only the generic case is considered.
For convenience%
\begin{equation}
	\lambda_{5}\ttt = A_1 \cdot e^{-\delta \! \frac{B_1}{\omega_{\text{d}}}} \lambda_{5,\text{n}} \ttt %
\end{equation}
is reformulated with the normalized quantity
\begin{equation}\label{eq:equation_lambda_5n}
	\lambda_{5,\text{n}}\!\left(t - \frac{B_1}{\omega_{\text{d}}}\right) = e^{\delta \, t} \, \sin\!\left(\omega_\text{d} \, t\right) - C_1 \, \text{.}
\end{equation}
In view of \eqref{eq:PMP_input_required_from_Hamiltonian} it remains to compute the switching points $t_i\in\left[0,t_\text{f}\right]$, $i=\left\{2,\dots,n-1\right\}$, i.e., the zeros of $\lambda_{5,\text{n}}\ttt$. Note that the Theorem on $n$-intervals \cite{Feldbaum1953} does not apply here, since the system has conjugate complex poles. Therefore, the number of switching points is not a priory fixed. However, an upper bound $\hat t_\text{f}=\frac{a_{\text{max}}}{j_{\text{max}}}+\frac{\pi}{\omega_{\text{d}}}$ for $t_\text{f}$ results from the application of a $\ZV$-shaper \cite{SingerPhdThesis1989,Singhose1990}, which satisfies all constraints of the posed optimal control problem while not being a time-optimal transition. As a consequence, independently of the constant $C_1$ and the damping $\delta$, the number of switching points is bounded from above. As a consequence, the following result holds:
\begin{theorem}\label{thrm:optimality}
	The optimal control problem \eqref{eq:optim_ctrl_problem} possesses a solution $\left(\bm{x},u,t_\text{f}\right)$, satisfying the following properties
	\begin{subequations}
		\begin{align}
			t_\text{f}&\in\left[\frac{a_{\text{max}}}{j_{\text{max}}}, \frac{a_{\text{max}}}{j_{\text{max}}}+ \frac{\pi}{\omega_{\text{d}}}\right) \label{eq:interval_for_tf}\\
			u\ttt&=\sum_{i=1}^{n-1} (-1)^{i+\nu}(H(t\!-\!t_i)-H(t\!-\!t_{i+1}))j_{\text{max}},\quad t_n=t_\text{f} \label{eq:optimal_solution:input}\\
			0&\le n-2\le
			2\cdot\left\lceil\frac{a_{\text{max}}}{j_{\text{max}}}\frac{\omega_{\text{d}}}{2\pi}\right\rceil\label{eq:number_of_switings_in_jerk_segment} \, \text{.} %
		\end{align}
	\end{subequations}
	Moreover, with \eqref{eq:equation_lambda_5n}, the switching points $t_2,\dots,t_{n-1}$ correspond to the zeros of $\lambda_{5,\text{n}}\ttt$, and 
	\begin{equation}\label{eq:optimal_solution:nu}
		\nu=\begin{cases}
			0,&\text{if} ~\lambda_{5,\text{n}}(0)>0 \, \text{,} \\
			1,&\text{if} ~\lambda_{5,\text{n}}(0)<0 \, \text{.}
		\end{cases}
	\end{equation}
Finally, the analytical solutions for the system state variables to such an input are given according to \hyperref[App:slider_move_from_impulse]{Appendix~\ref*{App:slider_move_from_impulse}} and \hyperref[App:sys_answer_to_trajectories]{Appendix~\ref*{App:sys_answer_to_trajectories}}. 
\end{theorem}
The brackets $\left\lceil \cdot \right\rceil$ in \eqref{eq:number_of_switings_in_jerk_segment} correspond to a \textit{ceil} function (value rounded up to the nearest integer) and define the upper limit of the number of switchings. To visualize this, two exemplary solutions for $\lambda_\text{5,n}\ttt$ are shown in \autoref{fig:exemplary_solution_lambda1_and_5}, which lead to exactly two switches at $t_2$ and $t_3$ (for a total of $n=4$, if $t_0$ and $t_\text{f}$ are also included for both examples). The parameters are chosen according to \autoref{tab:table_listing_kinematic_contraints} and \autoref{tab:table_listing_machine_parameters}. Note that for the result shown at the bottom of \autoref{fig:exemplary_solution_lambda1_and_5}, $n=6$ switches would be possible according to \eqref{eq:number_of_switings_in_jerk_segment}, but the amount of damping means that the solution only requires a total of $n=4$. Up to now, the optimal control problem considered has been reduced to the computation of the remaining variables $C_1$, $B_1$ and $t_\text{f}$, determining the shape of $\lambda_\text{5,n}\ttt$ and thus the switching points.

\bigskip
\textbf{Outlining the structure of the algorithm:} A graphical representation of the jerk segments can be introduced to explain the algorithm. %
This representation, where the jerk segments are visualized in the complex plane is introduced in \autoref{SubSec:Introduction_graphical_approach}. Afterwards, the calculation of the switching times is given in \autoref{SubSec:calculate_required_times_for_Delta_t} (developed to consider \eqref{eq:PMP_input_required_from_Hamiltonian}). The algorithm required to compute the final time $t_\text{f}$ (resp.\ the final angle $\varphi_\text{f}$) is provided after that in \autoref{SubSec:Efficient_algorithm_numerically}. %
Subsequently, time optimality is discussed in \autoref{SubSec:time_optimality_of_jerk_segments} and the algorithm is concluded in \autoref{SubSec:Conclusion_algorithm_and_EdgeCase}. Since a graphical representation is used to calculate the results, most of the times are converted to angles by multiplying them with the damped angular eigenfrequency $\omega_\text{d}$ to determine the correct orientation in the complex plane. The plots mostly show the times however, because it is easier to read. To make the connection to the angles, the period of oscillation denoted by $t_\text{v}$ (which corresponds to an angle of $2\pi$) is marked in most plots.

\subsection{Introduction of graphical approach}\label{SubSec:Introduction_graphical_approach}
As mentioned previously, the conditions with regards to the base displacement $x\ttt$ from \eqref{eq:boundary_constraints_PMP_for_system} can be met by requiring $\ddot{x}\tft=0$. Note that, instead of the variables $B_1$, $C_1$ and $t_{\text{f}}$ from \eqref{eq:equation_lambda_5n}, the switching times $t_{2},\dots,t_{n-1}$ and $t_\text{f}$ cf. \autoref{fig:lambda_for_multiple_segments}) can be used inside the calculations directly. According to \eqref{eq:optimal_solution:input} the solution of the optimal control problem corresponds to a piecewise constant jerk. This equation can be rewritten as
\begin{equation}\label{eq:trajectory_jerk}
	z^{\left(3\right)}\ttt = \sum\limits_{i=1}^{n} a_i H\!\left(t-t_i\right),\quad i=1,\dots,n,\quad t_n=t_\text{f} \, \text{.}
\end{equation}
with the coefficients $a_1,\dots,a_n$ satisfying.
\begin{equation}\label{eq:sum_ai_equal_zero}
	\sum_{i=1}^{n} a_i = 0 \, \text{.}
\end{equation}

\begin{figure}
	\centering
	\def\svgwidth{0.95\linewidth}
	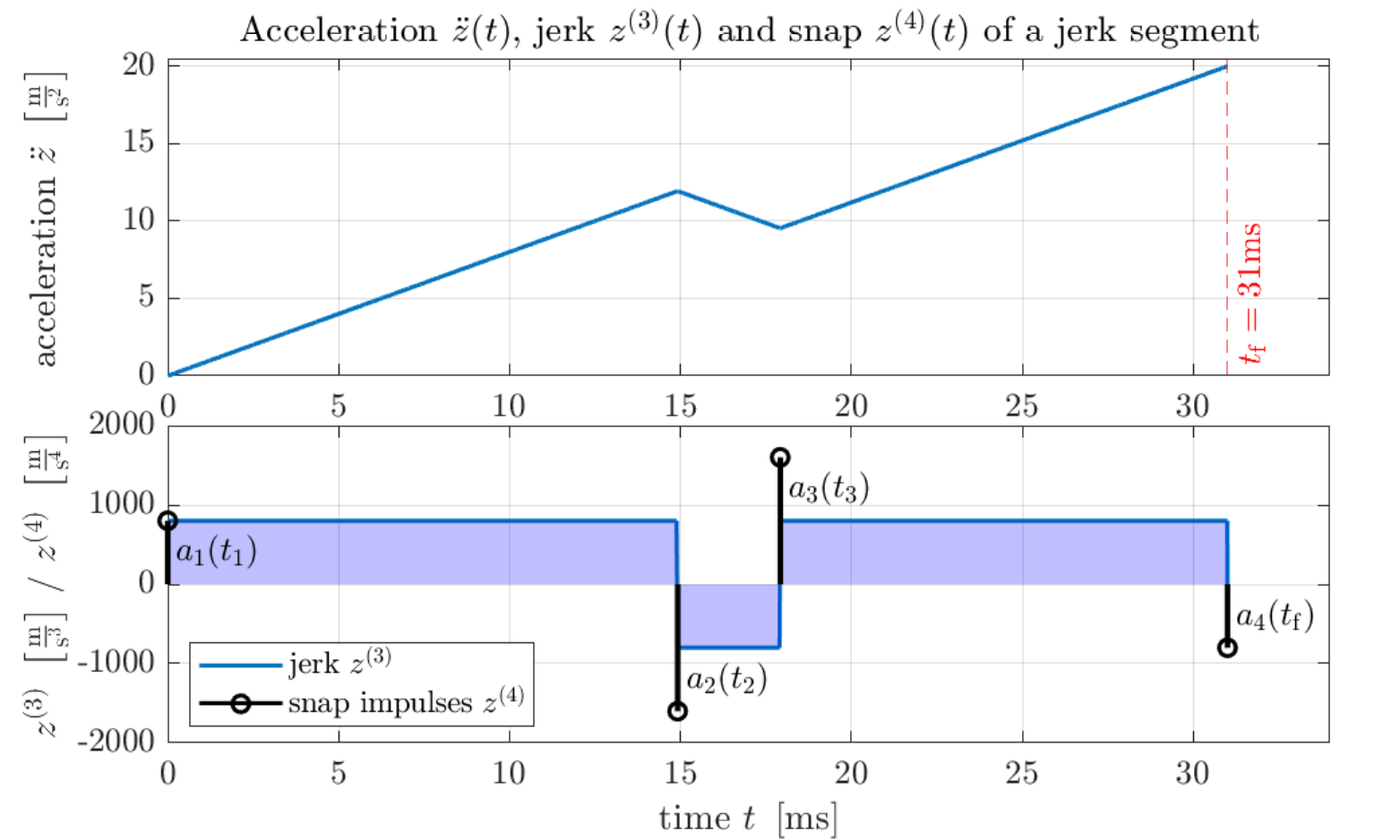 %
	\caption{Jerk segment to show the acceleration $\ddot{z}$, jerk $z^{\left(3\right)}$ and coefficients $a_1,\dots,a_4$.}
	\label{fig:OcpJ_segment_with_derivatives_zpp_zppp_zpppp}
\end{figure}
For the sake of illustration, an exemplary jerk segment, obtained for $a_\text{max} = \SI{20}{\meter\per\second\squared}$, is depicted in \autoref{fig:OcpJ_segment_with_derivatives_zpp_zppp_zpppp} to show the amplitudes $a_1,\dots,a_4$ of the respective steps %
as well as the acceleration $\ddot{z}$ and jerk $z^{\left(3\right)}$.

In order to determine the switching points, the trajectories $t\mapsto\dot{x}\ttt$ and
$t\mapsto\ddot{x}\ttt$ associated with the internal dynamics
are evaluated at $t=t_\text{f}$. For the piecewise constant input \eqref{eq:trajectory_jerk} these trajectories are given by \eqref{eq:xpp_sum_from_impulses} and \eqref{eq:xp_sum_from_impulses} in \hyperref[App:sys_answer_to_trajectories]{Appendix~\ref*{App:sys_answer_to_trajectories}}. Moreover, in view of \eqref{eq:sum_ai_equal_zero}, the expressions for $\dot{x}(t_\text{f})$ and $\ddot{x}(t_\text{f})$ simplify to
\begin{equation}\label{eq:equation_1and2_to_calc_t2t3tf}
	\begin{bmatrix}
		\dot{x}\tft \\ \ddot{x}\tft 
	\end{bmatrix} = \bm{0} = \bm{T} \bm{P} j_\text{max} \begin{bmatrix}
			\sum_{i=1}^{n=4}a_i^\star \, e^{\delta \, t_i} \, \cos\!\left(\omega_\text{d} \, t_i\right) \\
			\sum_{i=1}^{n=4}a_i^\star \, e^{\delta \, t_i} \, \sin\!\left(\omega_\text{d} \, t_i\right) 
	\end{bmatrix}
\end{equation}
with the matrices
\begin{subequations}
	\begin{align}
		\bm{T} &= \begin{bmatrix}
			\frac{m_\text{s}}{k} & \frac{m_\text{s} \, \delta}{m_\text{g} \, \omega_\text{d} \, \omega_0^2} \\
			0 & -\frac{m_\text{s}}{m_\text{g} \, \omega_\text{d}}
		\end{bmatrix} \, \text{,} \\
		\bm{P} &= \begin{bmatrix}
			e^{-\delta t_\text{f}} \, \cos\!\left(\omega_\text{d} \, t_\text{f}\right) & \hphantom{-}e^{-\delta t_\text{f}} \,\sin\!\left(\omega_\text{d} \, t_\text{f}\right) \\
			e^{-\delta t_\text{f}} \,\sin\!\left(\omega_\text{d} \, t_\text{f}\right) & -e^{-\delta t_\text{f}} \,\cos\!\left(\omega_\text{d} \, t_\text{f}\right)
		\end{bmatrix} \text{.}
	\end{align}
\end{subequations}
Therein, dependence on $t_\text{f}$ has been transferred to the matrix $\bm{P}$. The normalized coefficients $a_1^\star,\dots,a_n^\star$ defined by $a_i = a_i^\star j_\text{max}$
follow from \eqref{eq:optimal_solution:input} to be (cf.\ \autoref{fig:OcpJ_segment_with_derivatives_zpp_zppp_zpppp} for the case $n=4$)
\begin{equation}\label{eq:optimization:coeffs_sequences}
	a_i^\star=
	\begin{cases}
		(-1)^{\nu+1},&i=1\\\
		2 (-1)^{i+\nu},&i=2,\dots,n\!-\!1\\
		(-1)^{n+\nu},&i=n
	\end{cases}
\end{equation} 
Multiplying \eqref{eq:equation_1and2_to_calc_t2t3tf} with the inverse of the regular matrix $j_{\max}\bm{T}\bm{P}$ from the left and introducing the complex numbers ($i=1,\dots,n$)
\begin{equation}
	\begin{split}
		s_i&=a_i^\star e^{\delta \, t_i}\left(\cos\!\left(\omega_\text{d} \, t_i\right)+j\sin\!\left(\omega_\text{d} \, t_i\right)\right)=a_i^\star e^{(\delta+j\omega_\text{d}) \, t_i }\\
		 &=a_i^\star e^{(p_1+j)\varphi_i},
	\end{split}
\end{equation}
with the switching angles $\varphi_i = \omega_\text{d} t_i$ and the normalized damping parameter $p_1= \frac{\delta}{\omega_\text{d}}$, 
\eqref{eq:equation_1and2_to_calc_t2t3tf} can be equivalently rewritten as 
\begin{equation}\label{eq:graphical_problem_simplification_step_exp_complete}
	0 = \sum_{i=1}^{n} a_i^\star e^{(p_1+j) \varphi_i} = \sum_{i=1}^{n} s_i \, \text{.}
\end{equation}
A priory, of the switching times $t_1,\dots,t_n=t_\text{f}$, only $t_1 = 0$ is known.
The remaining points have to computed from the final conditions
\begin{equation}\label{eq:cal_aMax_from_impulses_SUM}
	a_\text{max} = \sum_{i=1}^{n}a_i t_i \, \Leftrightarrow
	a_{\text{max}}^\star:=\frac{\omega_\text{d} a_\text{max}}{j_{\text{max}}}=\sum_{i=1}^{n}a_i^\star\varphi_i
\end{equation}
in connection with \eqref{eq:graphical_problem_simplification_step_exp_complete} and the optimality condition \eqref{eq:equation_lambda_5n}.

\begin{figure}[!ht]
	\centering
	\def\svgwidth{0.95\linewidth}
	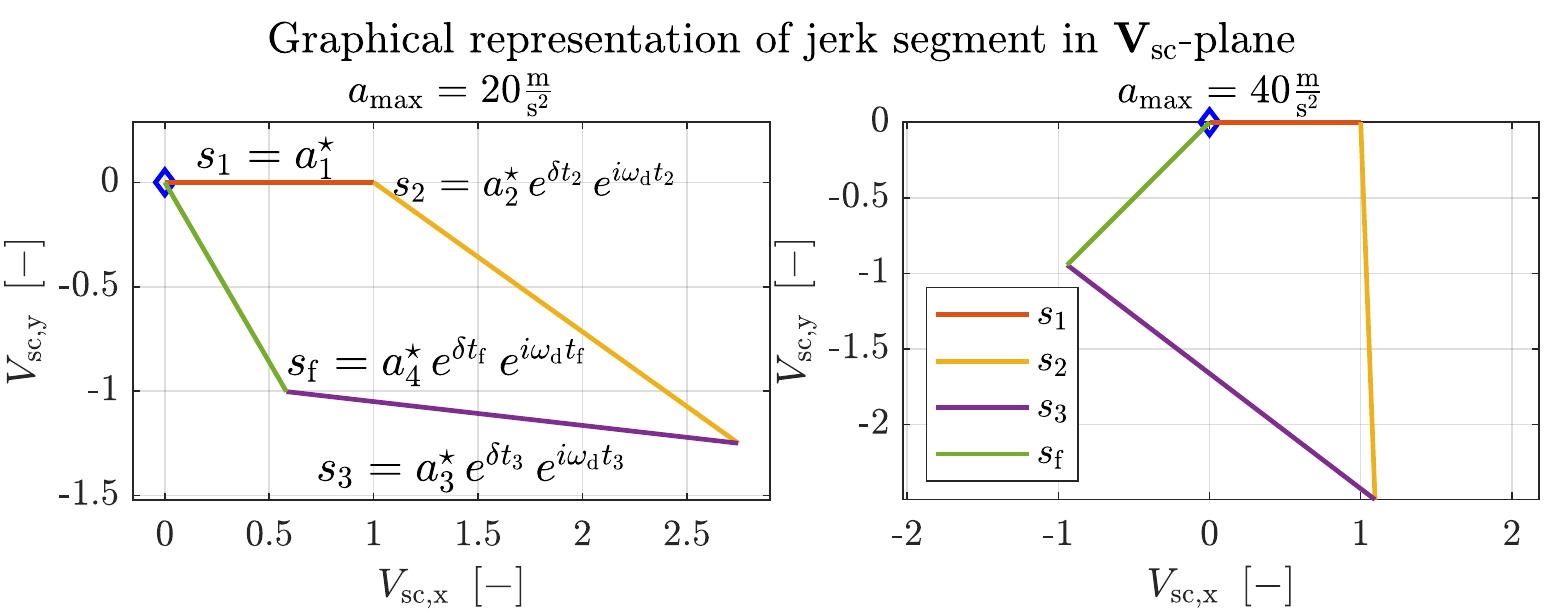 %
	\caption{Coefficients corresponding to the trajectories from \autoref{fig:exemplary_solution_lambda1_and_5} in the complex plane.}
	\label{fig:graphical_representation_to_solve_ocpJ}
\end{figure}

To visualize the graphical representation, the jerk segments from \autoref{fig:exemplary_solution_lambda1_and_5} are visualized according to \eqref{eq:graphical_problem_simplification_step_exp_complete} (both jerk segments with $n=4$) in \autoref{fig:graphical_representation_to_solve_ocpJ}.
For the first trajectory shown in the complex plane on the left-hand side of the plot (visualization of the top trajectory from \autoref{fig:exemplary_solution_lambda1_and_5}), the equations for calculating the vectors in the complex plane are also given directly in the plot for reference. Essentially, if all the parameters $a_i$ are concatenated as complex vectors $s_i$ using the times $t_i$ according to \eqref{eq:graphical_problem_simplification_step_exp_complete} and \eqref{eq:cal_aMax_from_impulses_SUM}, the result is a polygon in the complex plane (equations from \eqref{eq:equation_1and2_to_calc_t2t3tf} to calculate the real and imaginary parts). For the final jerk segment, this must be a closed polygon so that the oscillation of the base frame is zero when $a_\text{max}$ is reached, as can be seen in \eqref{eq:equation_1and2_to_calc_t2t3tf}. The bottom jerk segment of \autoref{fig:exemplary_solution_lambda1_and_5} is visualized on the right-hand side of \autoref{fig:graphical_representation_to_solve_ocpJ}.

\subsection{Calculating the switching-times of the jerk segment}\label{SubSec:calculate_required_times_for_Delta_t}
As seen in \autoref{fig:exemplary_solution_lambda1_and_5} one or multiple negative segments, i.e., segments with negative jerk, might be required to reach $a_\text{max}$ in a time-optimal fashion (depending on system damping). The switching times (or switching angles) depend on the terminal time. %
For a given transition time $t_\text{f}$, the overall duration of negative segments, called $\Delta t_\text{abs}$, can be calculated with \eqref{eq:calc_zpp_from_imp} from \hyperref[App:slider_move_from_impulse]{Appendix~\ref*{App:slider_move_from_impulse}} and \eqref{eq:boundary_constraints_PMP_for_system_zpp} from
\begin{equation}
	j_\text{max} t_\text{f} - 2 j_\text{max} \Delta t_\text{abs} = a_\text{max}
\end{equation}
to be
\begin{equation}\label{eq:switching_times:delta_tabs}
	\Delta t_\text{abs} = \frac{j_\text{max} t_\text{f} - a_\text{max}}{2 j_\text{max}} \, \text{.}
\end{equation}
When using the normalized quantities, this equation simplifies to
\begin{equation}\label{eq:switching_times:delta_phi_abs}
	\Delta\varphi_{\text{abs}}:=\omega_{\text{d}}\Delta t_{\text{abs}} = \frac{\varphi_\text{f} - a_{\text{max}}^\star }{2} \, \text{.}
\end{equation}
From \eqref{eq:switching_times:delta_phi_abs} and \eqref{eq:interval_for_tf} it follows immediately that $\varphi_\text{f} - a_{\text{max}}^\star\in[0,\pi)$ which implies 
\begin{equation}\label{eq:bounds_for_delta_phi_abs}
	0\le \Delta\varphi_{\text{abs}}<\frac{\pi}{2}.
\end{equation}
In this part of the algorithm, the switching times (resp.\ angles) will be computed up to a common but yet unknown shift. Those switching points are afterwards used in \autoref{SubSec:Efficient_algorithm_numerically} to compute the required shift and the corresponding switching times.
In the following, three cases are distinguished, the general case with positive damping and
the possibility of multiple negative jerk segments, the undamped case with $p_1=\delta=0$, and the case corresponding to one negative segment only. Naturally, the first of these cases is the most involved.

\begin{figure}[!ht]
	\centering
	\def\svgwidth{0.95\linewidth}
	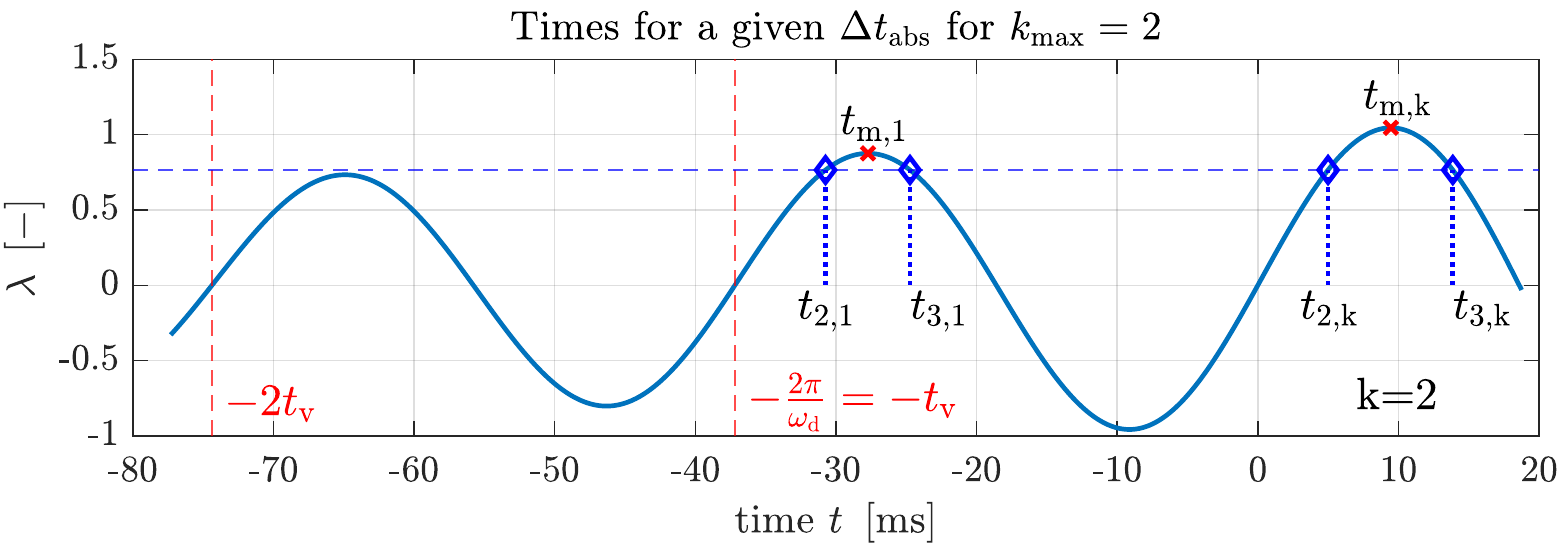 %
	\caption{Function $t\mapsto\lambda(\omega_\text{d} t)-C_1$ with switching times.}
	\label{fig:lambda_for_multiple_segments}
\end{figure}

\bigskip
\textbf{General case:} The case, where multiple segments are required only exists, if the damping of the system is sufficiently low. %
To illustrate the effect, the damping has been lowered to $10\%$ of the value from \autoref{tab:table_listing_machine_parameters} for the picture shown in \autoref{fig:lambda_for_multiple_segments}.
At this stage of analysis only the relative position of the switching angles w.r.t.\ $\varphi_{n-1}$ is of interest:
\begin{equation}\label{eq:relative_switching_times}
 	\Delta\varphi_i:=\varphi_{n-1}-\varphi_{2i}, \quad i=2,\dots,n-1.
\end{equation}
Therefore, the phase shift $B_1$ is neglected in \eqref{eq:equation_lambda_5n} which leads to
\begin{equation}\label{eq:general_shape_for_multiple_times}
	\lambda(\varphi):= \lambda_{5,\text{n}}\!\left(\tfrac{\varphi}{\omega_\text{d}}-\tfrac{B_1}{\omega_\text{d}}\right) =e^{p_1 \varphi} \sin\!\left(\varphi\right)-C_1 \, \text{.} %
\end{equation}
In the first step, the zeros of $\lambda$, i.e., the relative switching angles, are computed as a function of $C_1$. To this end, the local maxima of \eqref{eq:general_shape_for_multiple_times} are
determined. Differentiating \eqref{eq:general_shape_for_multiple_times} with respect to $\varphi$ gives
\begin{subequations}
	\begin{align}
		\lambda'(\varphi) &= e^{p_1 \varphi}\left(p_1 \sin\!\left(\varphi\right) + \cos\!\left(\varphi\right) \right) \\
		&= e^{p_1\varphi} \sqrt{p_1^2 + 1} \sin\!\left(\varphi + \Theta \right) \, \text{,} \\
		& \text{with} \ \Theta = \arccos \! \left(\frac{p_1}{\sqrt{p_1^2 + 1}}\right) \, \text{.}
	\end{align}
\end{subequations}
Hence, the maxima satisfy
\begin{equation}\label{eq:calculate_tm_where_peaks_are}
	\varphi_{\text{m},k} = {\left(2k+1\right) \pi - \Theta} \, \text{,} \quad k \in \mathbb{Z} \, \text{.}
\end{equation}

Although not rigorously proven in this contribution, it has turned out that only the case with $\nu=0$ and even $n$ is relevant, i.e., input trajectories that start and terminate with positive jerk. In this case, which solely is considered throughout the rest of the paper, it is convenient to introduce the number $\nel$ of maxima (or segments with negative jerk). Obviously, $n=2\nel+2$.

In the considered case, $C_1$ is always positive, as otherwise the condition \eqref{eq:switching_times:delta_phi_abs}, i.e.\ $\varphi_{\text{abs}}\le\frac{\pi}{2}$, would be violated.
Moreover, $\lambda$ possesses two zeros in the neighbourhood of each maximum $\varphi_{\text{m},k}$ satisfying $\lambda(\varphi_{\text{m},k})\ge 0$. More precisely, there is a zero
$\varphi_{2k}$ on $(\varphi_{\text{m},k}-\pi,\varphi_{\text{m},k})$ and a zero
$\varphi_{2k+1}$ on $(\varphi_{\text{m},k},\varphi_{\text{m},k}+\pi)$.
Having computed the maxima, the zeros $\varphi_{2},\dots,\varphi_{n-1}$ can be computed numerically since $\lambda$ is strictly monotonic on the respective intervals. This yields a function $\gamma$ defined by
\begin{align*}
	\gamma(C_1,\nel)&=(\gamma_2(C_1,\nel),\dots,\gamma_{2\nel+1}(C_1,\nel))\\
		&=(\varphi_2-\varphi_{2\nel+1},\dots,\varphi_{2\nel}-\varphi_{2\nel+1},0).
\end{align*}
For a given total transition time $\phif$ all relative switching angles $n$ must
be contained in the interval $[-\phif,0]$. Therefore,
\begin{equation}
	\nel=\eta(C_1,\phif)=\max\{\hat \nel\in\mathbb{N}^+|\gamma_2(C_1,\hat\nel)+\phif>0\}.
\end{equation}
Afterwards, $\Delta\phiabs$ can be computed by
\begin{align*}
	\Delta\phiabs=\rho(C_1,\eta(\phif)):=\sum_{i=1}^{\nel}(\varphi_{2i+1}-\varphi_{2i})
\end{align*}
with $\varphi_i=\gamma_2(C_1,\eta(\phif))$.
With \eqref{eq:switching_times:delta_phi_abs}, i.e., $\Delta\phiabs=\frac{\phif-\amax^\star}2$
it follows:
\begin{align*}
	\frac{\phif-\amax^\star}{2}=\rho(C_1,\eta(\phif)).
\end{align*}
This latter relation can be solved for $C_1$ for $\phif$ from the interval $[\amax^\star,\amax^\star+\pi]$ which in turn delivers required relative switching angles by evaluating $\gamma(C_1,\eta)$.
This part of the algorithm is the computationally most expensive one within the motion planning scheme. It finally provides a relation between the total transition time $\phif$ and
the relative switching angles $\varphi_2,\dots,\varphi_{2\nel+1}$ where the number $\nel$ of negative jerk segments is another result of the algorithm. This relation is required in subsequent calculations in \autoref{SubSec:Efficient_algorithm_numerically}. Therefore, in order to allow for a fast execution, it is convenient to precompute this map once for several points from $[\amax^*,\amax^*+\pi]$ and afterwards interpolate between those points.

\begin{figure}[!ht]
	\centering
	\def\svgwidth{0.95\linewidth}
	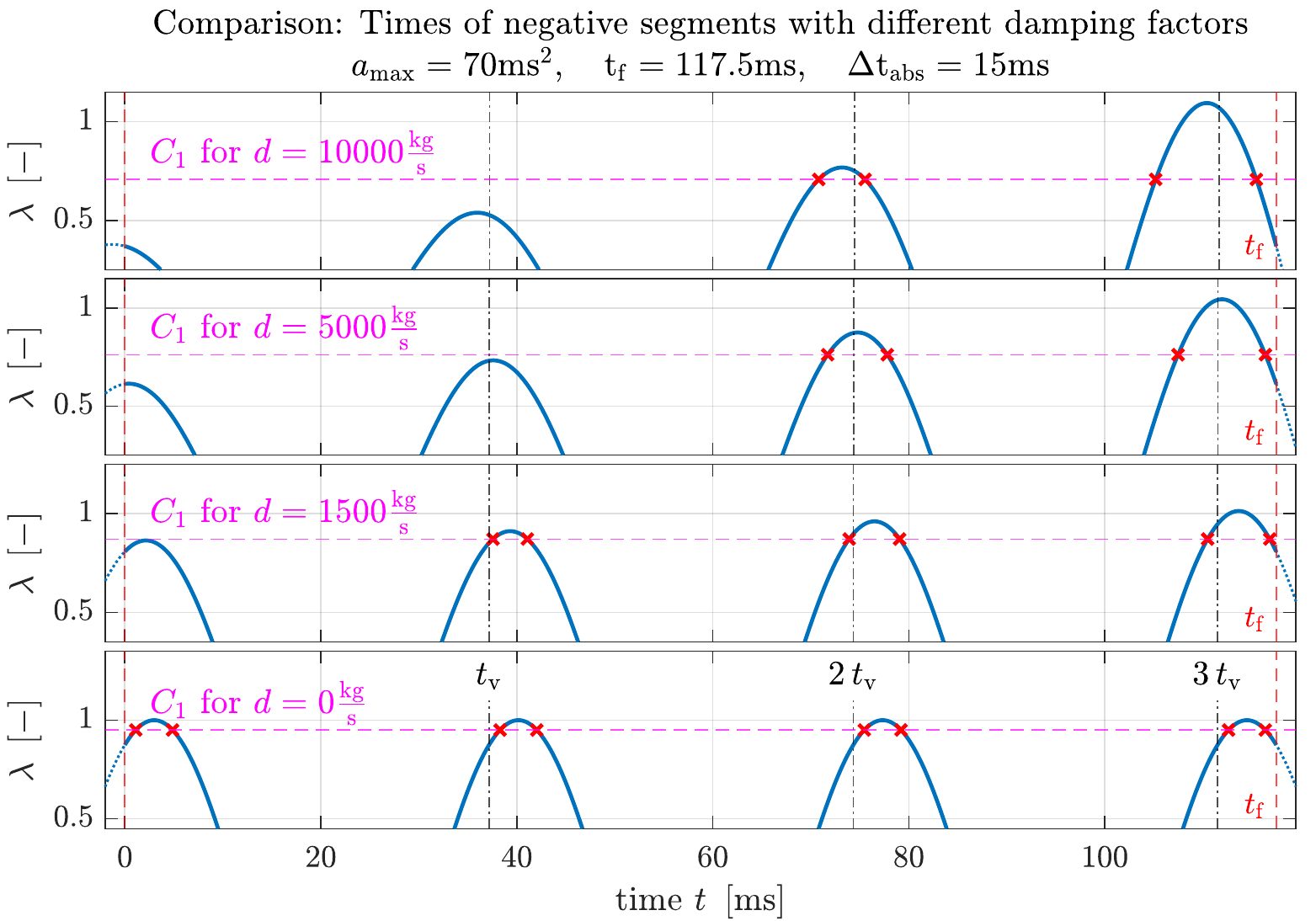 %
	\caption{Graphical representation of all cases with different values for the damping parameters, where $t_\text{f}$ and $\Delta t_\text{abs}$ coincide.}
	\label{fig:comparison_times_all_cases}
\end{figure}

\bigskip
\textbf{Zero damping:}
If the damping equals zero, every negative segment of the solution has the same width. %
Therefore, the number of negative segments directly depends on the time $t_\text{f}$ and can be calculated with
\begin{equation}
	n_\text{el} = \left\lceil\frac{\omega_{\text{d}} t_\text{f}}{2\pi}\right\rceil=\left\lceil\frac{\phif}{2\pi}\right\rceil \, \text{.}
\end{equation}
By a symmetry argument (resulting from $\delta = 0$, cf. \eqref{eq:equation_lambda_5n}), it follows that ($k=1,\dots,n_\text{el}$)
	\begin{align}
		\Delta\varphi_{2k} &= 2 \pi(n_\text{el} - k) + \frac{\Delta \varphi_\text{abs}}{n_\text{el}},&%
		\Delta\varphi_{2k+1} &= 2 \pi(n_\text{el} - k ). %
	\end{align}
A jerk segment corresponding to $d=0$ is visualized in \autoref{fig:comparison_times_all_cases}.

\bigskip
\textbf{One negative jerk segment:}
If there is only one negative segment, i.e.\ $\nel=1$,
 \eqref{eq:switching_times:delta_phi_abs} reduces to 
\begin{equation}
	\Delta \varphi_{\text{abs}}=\varphi_3-\varphi_2 = \frac{\varphi_\text{f}-\amax^\star}{2},
\end{equation}
Hence, according to \eqref{eq:relative_switching_times}, 
\begin{equation}
	\Delta\varphi_{2}= \varphi_3-\varphi_2,\quad \Delta \varphi_{3}=0\, \text{.}
\end{equation}
This case is shown on top in \autoref{fig:exemplary_solution_lambda1_and_5} and, moreover in \autoref{fig:OcpJ_segment_with_derivatives_zpp_zppp_zpppp}.

\bigskip
\textbf{Graphical representation of results:}
As mentioned previously all the cases are shown in \autoref{fig:comparison_times_all_cases} for different values of $\delta$, but with the same $t_\text{f}$ and $\Delta t_\text{abs}$. Those are only meant to show, how $\Delta t_\text{abs}$ is split up between multiple negative segments (where required), relative to each other for different values of $\delta$. For this reason, only the constraints with respect to the slider motion governed by \eqref{eq:BC_ztzt_l3tft_segment}, \eqref{eq:BC_zptzt_l4tft_segment} and \eqref{eq:boundary_constraints_PMP_for_system_zpp} are considered here. All of the curves are shifted in time axis in a way, where $\bm{\ell}_\text{m1}$ and $\bm{\ell}_\text{m2}$ are parallel, as shown in \autoref{fig:graphical_representation_ocpJ_three_cases_for_tf}. Actual solutions for those parameters, where all of the constraints related to base oscillation, meaning \eqref{eq:BC_xtzt_xtft_segment} and \eqref{eq:BC_xptzt_xptft_segment} are taken into account as well, are shown at the end of this chapter in \autoref{fig:finished_jerkTrajectory_segments_four_parameters_accel_jerk}. The curves given here are only meant to show the influence of different damping parameters on the number of negative segments as an illustrative example.
The period $t_\text{v}$ of oscillation are also marked in the plots. Those times are slightly different since varying damping ratios lead to slightly different damped eigenfrequencies. However, these small differences are not visible with time axis scaling used. Depending on the damping, more than one negative segment might be required as visible in the plot. The calculation of the critical damping, when more than one negative segment is required is discussed in \autoref{SubSec:time_optimality_of_jerk_segments}.

\subsection{Algorithm to calculate the jerk segments numerically}\label{SubSec:Efficient_algorithm_numerically}
\begin{figure}[!ht]
	\centering
	\def\svgwidth{0.95\linewidth}
	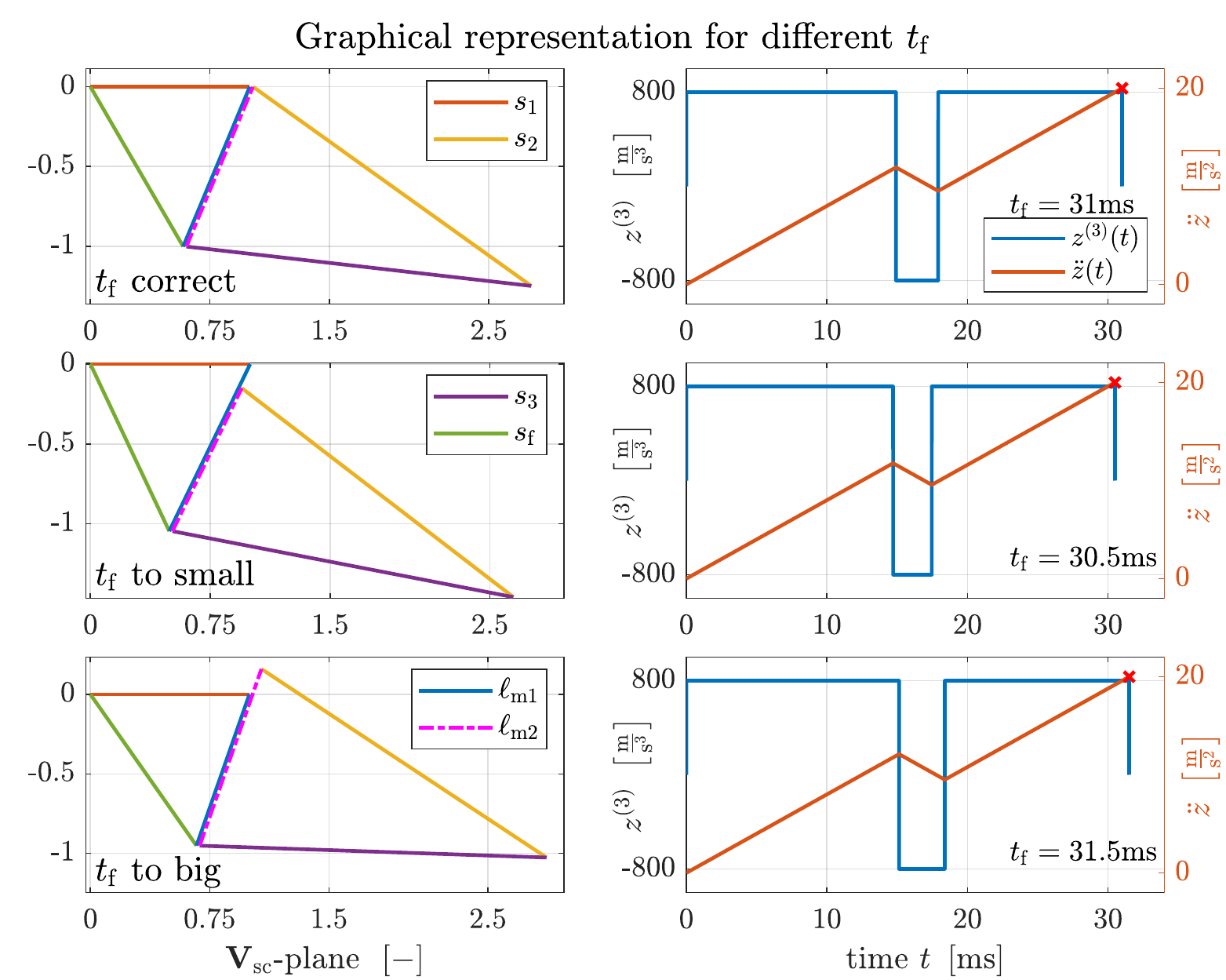 %
	\caption{Matching the absolute values of $\vellm1$ and $\vellm2$ by varying $\varphi_\text{f}$ (respectively $t_\text{f}$).}
	\label{fig:graphical_representation_ocpJ_three_cases_for_tf}
\end{figure}

Based on the results of the previous subsection, this section introduces the calculation required to solve the problem efficiently for any number of switching points. The problem is finally reduced to a line search problem with only one free variable $\varphi_\text{f}$ to reduce the complexity of the calculation. The variation of $\varphi_\text{f}$ then leads to the solution as shown in \autoref{fig:graphical_representation_to_solve_ocpJ}. The boundaries, of $\varphi_\text{f}$ are known and given in \eqref{eq:interval_for_tf}. Up to a common shift, the angles $\varphi_{2,k}=\varphi_{2k}$ and $\varphi_{3,k}=\varphi_{2k+1}$ can be calculated directly from $\varphi_\text{f}$, as given by \autoref{SubSec:calculate_required_times_for_Delta_t}. When varying $\varphi_\text{f}$, the three cases shown in \autoref{fig:graphical_representation_ocpJ_three_cases_for_tf} can occur.
Those have been altered already to fit the solution method explained in the following. Equation \eqref{eq:graphical_problem_simplification_step_exp_complete} can be rewritten as:
\begin{equation}\label{eq:numerics:matching_condition}
	\underbrace{s_1+s_n}_{\bm{\ell}_{\text{m1}}(\phif)}=\underbrace{-\sum_{i=2}^{n-1}s_i}_{\bm{\ell}_{\text{m2}}(\phif,\varphi_\text{f})},\quad s_i=a_i^\star e^{(p_1+j)\varphi_i}
\end{equation}
The left hand side of this equation spans a triangle in the complex plane while the right hand side corresponds to a polygon with $n-1$ edges (cf.\ \autoref{fig:graphical_representation_ocpJ_three_cases_for_tf} for $n=4$ and \autoref{fig:finished_jerkTrajectory_segments_four_parameters_xyPlane} for $n\in\{4,6,8\}$). For a chosen sequence of coefficients, %
the left hand side of \eqref{eq:numerics:matching_condition} is completely determined by
$\phif$, i.e.,
\begin{equation}
	\bm{\ell}_{\text{m1}}(\phif)=s_1+s_n=a_1^\star+a_n^\star e^{(p_1+j)\phif}.
\end{equation}
Moreover, as
\begin{equation}\label{eq:numerics:phi_from_deltaphi}
	\varphi_i=\varphi_{n-1}-\Delta\varphi_i, \quad i=2,\dots,n-1
\end{equation}
where, according to \autoref{SubSec:calculate_required_times_for_Delta_t} $\Delta\varphi_i$ is determined by $\phif$, the right hand side of \eqref{eq:numerics:matching_condition} can be rewritten in the form
\begin{equation}
	\bm{\ell}_{\text{m2}}(\phif,\varphi_{n-1})=-\sum_{i=2}^{n-1}s_i=e^{(p_1+j)\varphi_{n-1}}\bar{\bm{\ell}}_{\text{m2}}(\phif)
\end{equation}
with
\begin{equation}
\bar{\bm{\ell}}_{\text{m2}}(\phif)=-\sum_{i=2}^{n-1} a_i^\star e^{-(p_1+j)\Delta\varphi_i}.
\end{equation}
As a consequence, \eqref{eq:numerics:matching_condition} appears in the form
\begin{equation}
\bm{\ell}_{\text{m1}}(\phif)= e^{(p_1+j)\varphi_{n-1}}\bar{\bm{\ell}}_{\text{m2}}(\phif).
\end{equation}
The argument of the right hand side can be matched to that of the left hand side by computing
\begin{equation}
	\bar{\psi}(\phif)=\arg\!\left(\bm{\ell}_{\text{m1}}\!\left(\phif\right)\right)-\arg\!\left(\bar{\bm{\ell}}_{\text{m2}}\!\left(\phif\right)\right)
\end{equation}
This step matches the angles of $\bm{\ell}_{\text{m1}}\!\left(\phif\right)$ and $\bar{\bm{\ell}}_{\text{m2}}\!\left(\phif\right)$ to be parallel, as shown in \autoref{fig:graphical_representation_ocpJ_three_cases_for_tf}.

To ensure compliance with the optimality conditions (see also \autoref{fig:exemplary_solution_lambda1_and_5} and \autoref{fig:lambda_for_multiple_segments}), the negative part of the jerk segments must be placed as far as possible towards $t_\text{f}$ (while still remaining smaller than $t_\text{f}$). This is achieved by choosing
\begin{equation}\label{eq:calc_phi_3n_fixed_with_2Pi}
	\varphi_{n-1}=\psi(\phif)=\bar{\psi}(\phif)+2 \pi\left\lfloor\frac{\varphi_\text{f} -\psi(\phif)}{2 \pi}\right\rfloor.
\end{equation}
where $\left\lfloor  \cdot \right\rfloor$ represents the \textit{floor} function (rounding down to the nearest integer). As the argument of a complex number is defined only up to a multiple of $2\pi$, the latter correction is required to comply with the switching law \eqref{eq:PMP_input_required_from_Hamiltonian} in view of the periodic switching function \eqref{eq:equation_lambda_5n}. This ensures that the final switching occurs at the zero of \eqref{eq:equation_lambda_5n} closest to $\phif$.

Having matched the arguments of $\bm{\ell}_{\text{m1}}$ and $\bm{\ell}_{\text{m2}}$, the problem is reduced to the computation of the zeros of 
\begin{equation}\label{eq:error_depending_on_phiF}
	\bar{\ell}\!\left(\phif\right):=\left|\bm{\ell}_{\text{m1}}(\phif)\right|^2-\left|e^{(p_1+j)\psi(\phif)}\bar{\bm{\ell}}_{\text{m2}}(\phif)\right|^2 \, \text{.}
\end{equation}
This step matches the lengths of $\bm{\ell}_{\text{m1}}\!\left(\phif\right)$ and ${\bm{\ell}}_{\text{m2}}\!\left(\phif\right)$ to be equal, as shown in \autoref{fig:graphical_representation_ocpJ_three_cases_for_tf}. Note, that the lengths are matched through variation of $\varphi_\text{f}$. This is achieved numerically by a line search on the interval $\left[a_\text{max}^\star,a_\text{max}^\star+\pi\right]$ (cf.\ \eqref{eq:interval_for_tf}). Here $\varphi_\text{f}$ is adjusted based on the value of $\bar{\ell}\!\left(\phif\right)$.

\begin{figure}[]
	\centering
	\def\svgwidth{0.95\linewidth}
	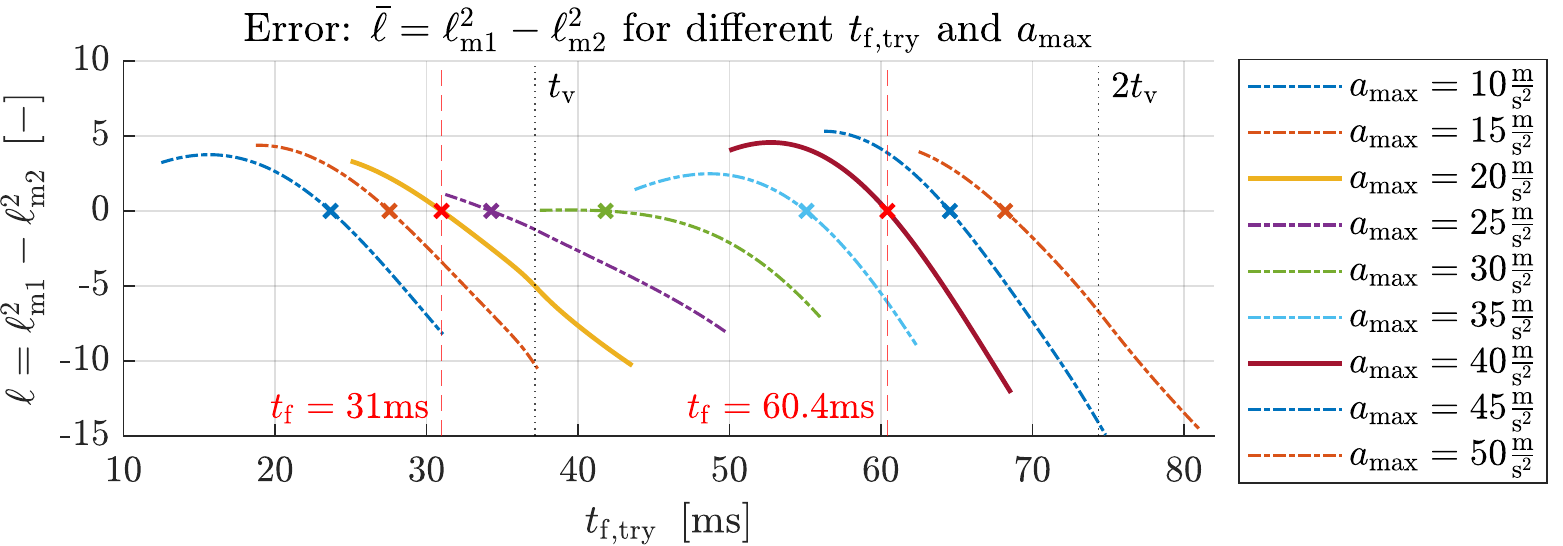 %
	\caption{Showing $\bar{\ell}$ and the final $t_\text{f}$ for jerk segments of different acceleration.}
	\label{fig:error_between_the_lengths_on_interval}
\end{figure}

The errors $\phif\mapsto\bar{\ell}\!\left(\phif\right)$ for different maximal accelerations $a_\text{max}$ are shown on the entire intervals \eqref{eq:interval_for_tf} in \autoref{fig:error_between_the_lengths_on_interval}. The jerk segments corresponding to the trajectories in \autoref{fig:exemplary_solution_lambda1_and_5} are marked by the solid lines. 
After $\varphi_\text{f}$ is calculated, the angle $\varphi_{n-1}$ can be computed directly by using \eqref{eq:calc_phi_3n_fixed_with_2Pi} which in turn gives $\varphi_{2},\dots,\varphi_{n-2}$ by \eqref{eq:numerics:phi_from_deltaphi}. The switching times follow immediately from the switching angles via $t_i=\frac{\varphi_i}{\omega_\text{d}}$. The implementation chosen for use on a PLC is a binary search (for its numerical robustness), as described in \autoref{alg:Calc_jerk_segment}.
\begin{algorithm}
	\caption{Line search to calculate jerk segment}\label{alg:Calc_jerk_segment}
	\begin{algorithmic}[1]
		\State Initialize $\varphi_\text{span} = \pi$ \Comment{See \eqref{eq:interval_for_tf}}
		\State Set $\varphi_\text{f,min} = a_{\text{max}}^\star = \frac{a_\text{max}}{j_\text{max}} \cdot \omega_{\text{d}}$ \Comment{See \eqref{eq:interval_for_tf} and \eqref{eq:cal_aMax_from_impulses_SUM}}
		\State Set $\varphi_\text{f,try} = \varphi_\text{f,min} + \nicefrac{1}{2}\cdot \varphi_\text{span}$
		
		\For{$n_\text{iter} = 2 \ \text{to} \ n_\text{iter,max}$}
		\State Calculate $\bar{\ell}\!\left(\varphi_\text{f,try}\right)$ with \eqref{eq:error_depending_on_phiF}
		\If{$\bar{\ell}\!\left(\varphi_\text{f,try}\right) < 0$}
		\State $\varphi_\text{f,try} = \varphi_\text{f,try} - \left(\frac{1}{2}\right)^{n_\text{iter}} \cdot \varphi_\text{span}$
		\Else
		\State $\varphi_\text{f,try} = \varphi_\text{f,try} + \left(\frac{1}{2}\right)^{n_\text{iter}} \cdot \varphi_\text{span}$
		\EndIf
		\EndFor
		\State Calculate times $t_i = \frac{\varphi_i}{\omega_{\text{d}}}$ from $\varphi_i$ of last iteration
		\State Get coefficients $a_i$ according to \eqref{eq:optimization:coeffs_sequences}
		\State Calculation of jerk segment parameters $t_i$ and $a_i$ finished
	\end{algorithmic}
\end{algorithm}
This method of calculating jerk segments was implemented on the laboratory setup shown in \autoref{sec:measurements_lab_sys}, where all calculations are implemented in single precision floating point format. It was found that after 18 iterations $\left(n_\text{iter,max} = 18\right)$ the value for $\varphi_\text{f,try}$ does not change due to the single precision of the binary refinement and the addition with $\varphi_\text{f,min}$. As far as computational complexity is concerned, this calculation leads to exactly the same calculation procedure with the same number of steps, independent of the parameters. As a reference, on the PLC used on the laboratory system (X20CP1585: see \autoref{sec:measurements_lab_sys}), the time to compute a single jerk segment measured on the PLC is $t_\text{CPU} = \SI{326}{\micro\second}$ (average of 10 measurements with $t_\text{CPU,min} = \SI{324}{\micro\second}$ and $t_\text{CPU,max} = \SI{328}{\micro\second}$). This method was chosen over other methods due to its numerical robustness.

\subsection{Discussion of time-optimality of jerk segments}\label{SubSec:time_optimality_of_jerk_segments}
\begin{figure}[!ht]
	\centering
	\def\svgwidth{0.95\linewidth}
	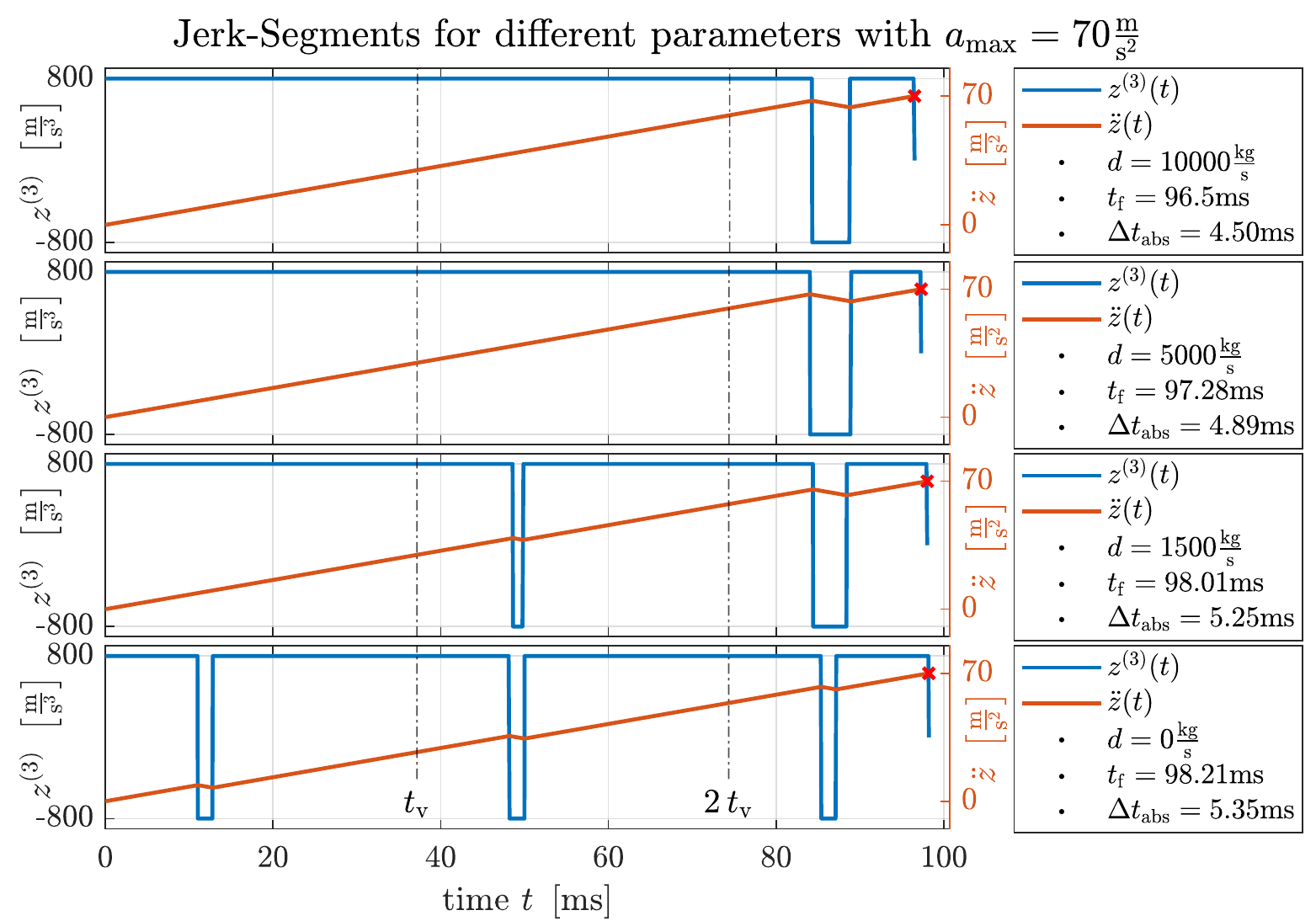 %
	\caption{Actual jerk segments for the same parameters as provided by \autoref{fig:comparison_times_all_cases}.}
	\label{fig:finished_jerkTrajectory_segments_four_parameters_accel_jerk}
\end{figure}

\begin{figure}[!ht]
	\centering
	\def\svgwidth{0.95\linewidth}
	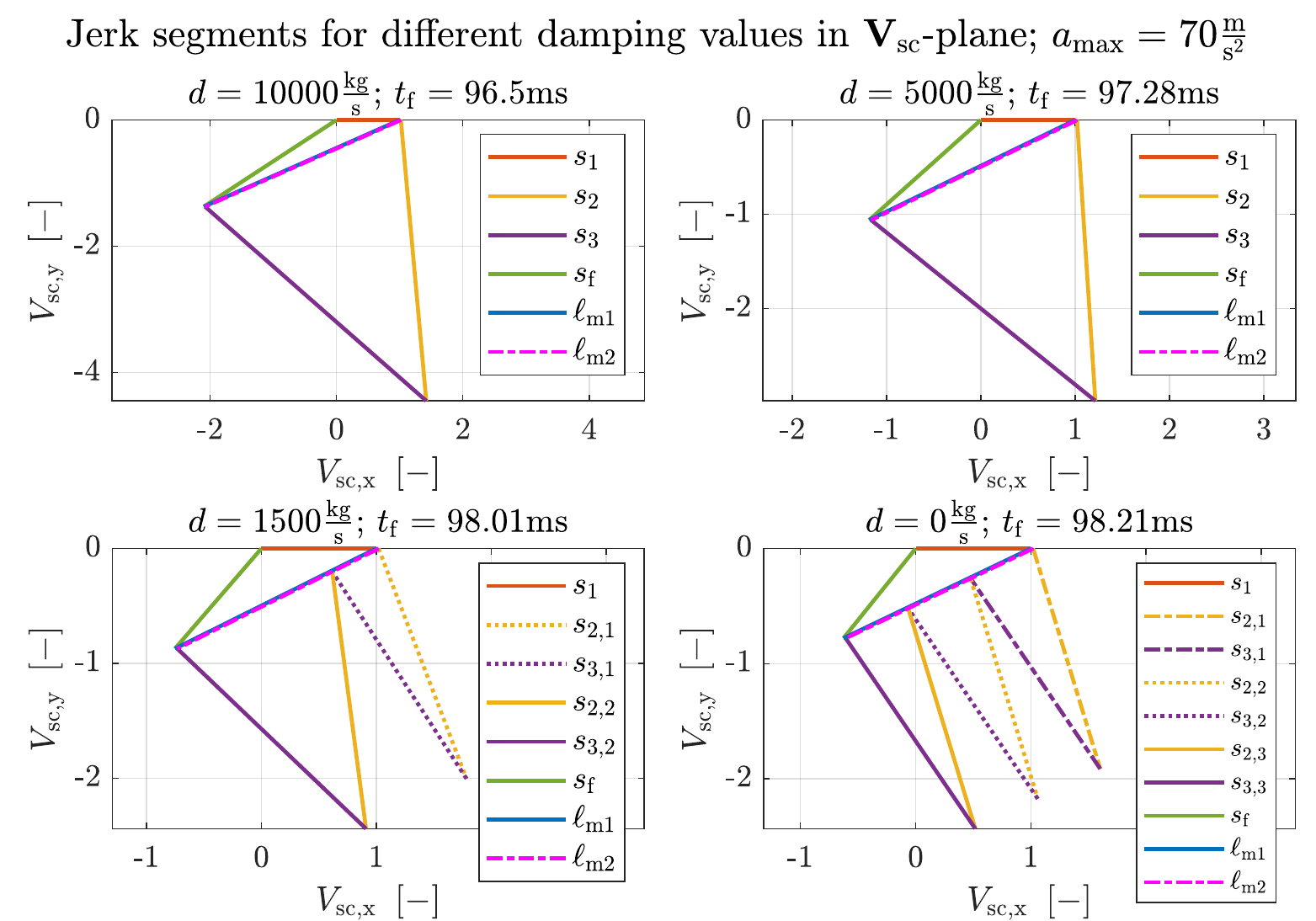 %
	\caption{Actual jerk segments for the same parameters as provided by \autoref{fig:comparison_times_all_cases}. Trajectories from \autoref{fig:finished_jerkTrajectory_segments_four_parameters_accel_jerk} represented in the complex plane. Effect on damping best visible when comparing the length's of $\bm{s}_1$ with $\bm{s}_\text{f}$.}
	\label{fig:finished_jerkTrajectory_segments_four_parameters_xyPlane}
\end{figure}

\textbf{Full jerk segments as calculated with the algorithm:}
The presented algorithm is applied to the same parameters that were used in \autoref{fig:comparison_times_all_cases} to show the computed jerk segments that satisfy all the constraints provided by \eqref{eq:startState_system_ocpJ_jerkSegment} and \eqref{eq:endState_system_ocpJ_jerkSegment}. The corresponding trajectories are shown in \autoref{fig:finished_jerkTrajectory_segments_four_parameters_accel_jerk}.
The required values for $\Delta t_\text{abs}$ are given and the resulting $t_\text{f}$ as well. %
The representation of those jerk segments in the complex plane is shown in \autoref{fig:finished_jerkTrajectory_segments_four_parameters_xyPlane}. The representation of the jerk segments from \autoref{fig:finished_jerkTrajectory_segments_four_parameters_accel_jerk} in \autoref{fig:finished_jerkTrajectory_segments_four_parameters_xyPlane} is the same as the visualization of \autoref{fig:exemplary_solution_lambda1_and_5} in \autoref{fig:graphical_representation_to_solve_ocpJ} (see above for a detailed explanation).
All of those jerk segments comply with the necessary conditions for a local optimum as outlined by the PMP. Since, this is the only possible solution that allows to satisfy the constraints \eqref{eq:boundary_constraints_PMP_for_system} while complying with the switching law for the input \eqref{eq:PMP_input_required_from_Hamiltonian}, it can be concluded, that this is in fact the globally best solution for the problem defined.

\begin{figure}[!ht]
	\centering
	\def\svgwidth{0.95\linewidth}
	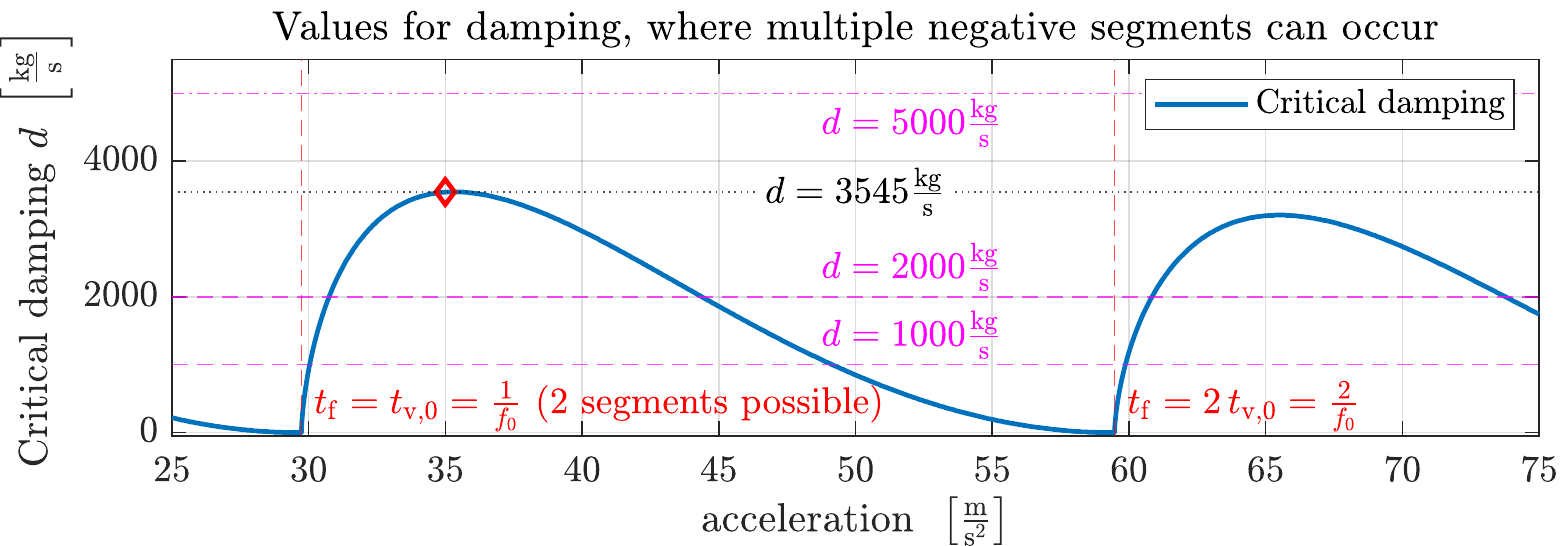 %
	\caption{Values for damping, when multiple segments are required. Compared to the damping from \autoref{tab:table_listing_kinematic_contraints}, the damping where multiple negative segments can actually occur is lower.}
	\label{fig:analysis_critical_damping}
\end{figure}

\bigskip
\textbf{Analysis of critical damping:}
This section continues with further analysis regarding the number of segments and its connection to time optimality. As shown in \autoref{fig:comparison_times_all_cases}, depending on the damping of the system, multiple negative segments might be required to ensure optimality of the jerk segments. Analysis showing the critical damping for different $a_\text{max}$ values is shown in \autoref{fig:analysis_critical_damping}. If the magnitude of the damping parameter is below this value, more than one negative section is required in the jerk segment.
To obtain the critical damping, starting from an initial value, the damping has been decreased until more than one negative segment was required for the computation of the jerk segments. This procedure has been performed for different values of $a_\text{max}$. If the system damping is higher than the critical damping, which is the case for the parameters in \autoref{tab:table_listing_machine_parameters}, the solution with one negative segment is the time-optimal solution for the jerk segments. It can be seen, that the critical damping depends on the maximal acceleration $a_\text{max}$
(also compare the jerk segment for $a_\text{max}=\SI{29}{\meter\per\second\squared}$ and $a_\text{max}=\SI{36}{\meter\per\second\squared}$ shown as subplots in \autoref{fig:showing_the_edge_case_occuring}).
\begin{figure}[!ht]
	\centering
	\def\svgwidth{0.95\linewidth}
	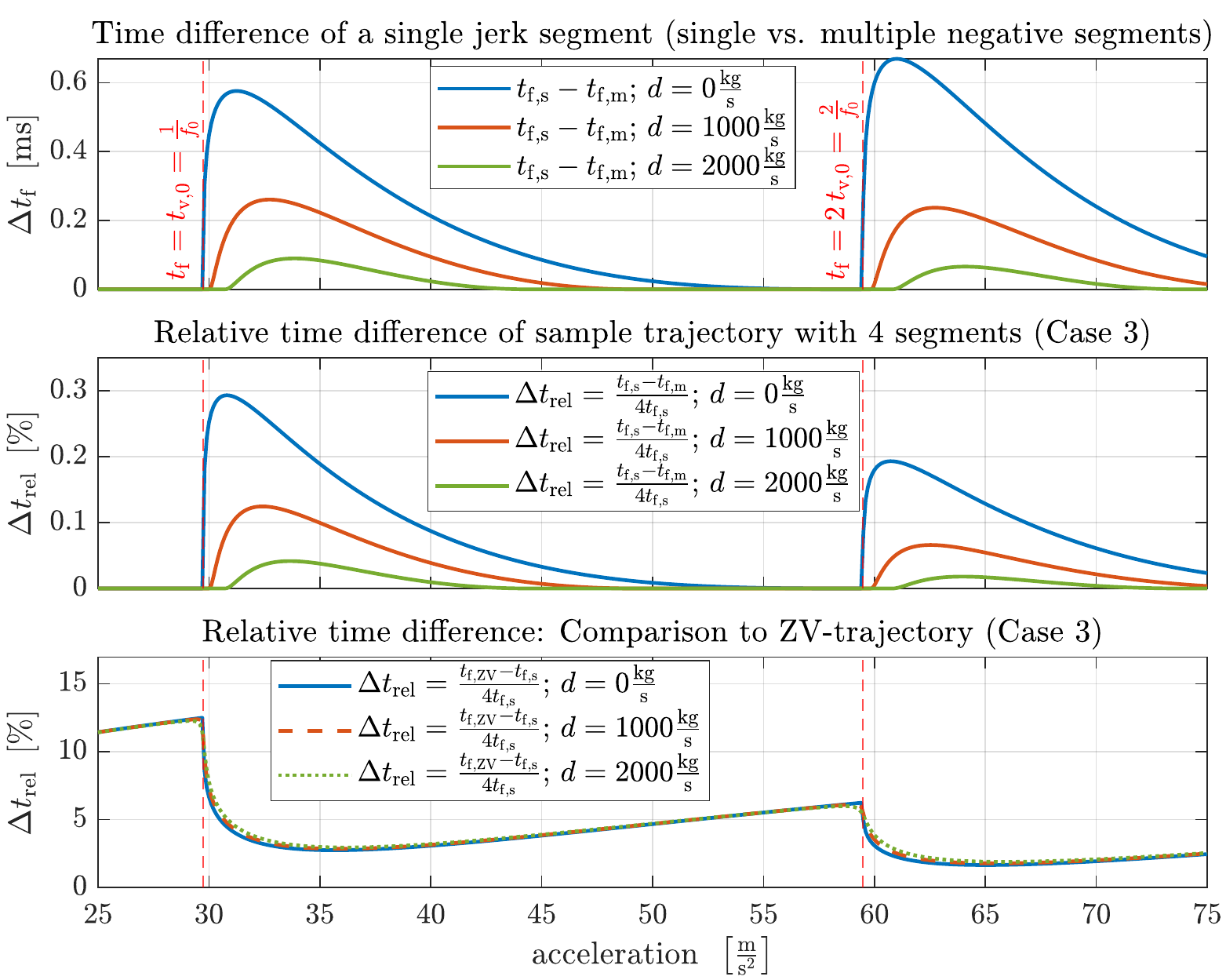
	\caption{Actual time-advantage of jerk segments, when using more than one negative section \autoref{fig:comparison_times_all_cases}}%
	\label{fig:analysis_timeAdvantage_multiple_neg_segments}
\end{figure}
Subsequently, the case of multiple vs. a single negative section is further discussed. The absolute time gain of using multiple negative sections versus a single one for three different damping values is shown in \autoref{fig:analysis_timeAdvantage_multiple_neg_segments} (top subplot). Throughout this plot the damping values used for the comparison have to be different from the one given in \autoref{tab:table_listing_machine_parameters}, as this set of parameters requires exactly one negative section, regardless of the maximum acceleration (shown in \autoref{fig:analysis_critical_damping}). Since a point-to-point movement consists of multiple jerk segments, the middle plot shows a comparison of the relative time advantage when four such jerk segments are assembled to a full trajectory. Notice that, as the acceleration $a_\text{max}$ and the duration $t_\text{f}$ of the jerk segments increase, the relative time advantage decreases. For the particular case of a system without damping, the time advantage when comparing two such $\ocpJ$ trajectories (one with a single negative section in its jerk segments and one with multiple) is as low as $0.3\%$. In contrast, the time advantage of the $\ocpJ$ approach can be as high as $12\%$, when comparing to a $\ZV$-shaped trajectory. For very high accelerations (other relevant extreme case), the time $t_\text{f}$ to reach those accelerations naturally increases and in the case of a system with damping, the oscillation caused by the first impulse at $t=0$ can naturally decay further. This means that the time required for the negative parts of the jerk segments will also decrease, reducing the chance that more negative parts will be required. At the same time, the relative time advantage of using multiple negative segments decreases, as \autoref{fig:analysis_timeAdvantage_multiple_neg_segments} shows. A detailed analysis of the time advantage of the $\ocpJ$ approach compared to other path planning approaches is a major contribution of \cite{tau_ocpJ_assembly_part1}.
To summarize for the jerk segments, the time advantage of taking multiple negative sections is smaller than $0.3\%$. %
Based on this analysis, the following conjecture can be formulated:
\begin{conjecture}
	Even if the optimal solution requires more than one negative section within the jerk segment, there are trajectories with only one negative segment which comply with the terminal constraints. Moreover, the numerical results presented in \autoref{fig:analysis_timeAdvantage_multiple_neg_segments} show, that the actual time advantage of using multiple negative sections is small, if not negligible.
\end{conjecture}

\subsection{Edge-case}\label{SubSec:Conclusion_algorithm_and_EdgeCase}
\begin{figure}[!ht]
	\centering
	\def\svgwidth{0.95\linewidth}
	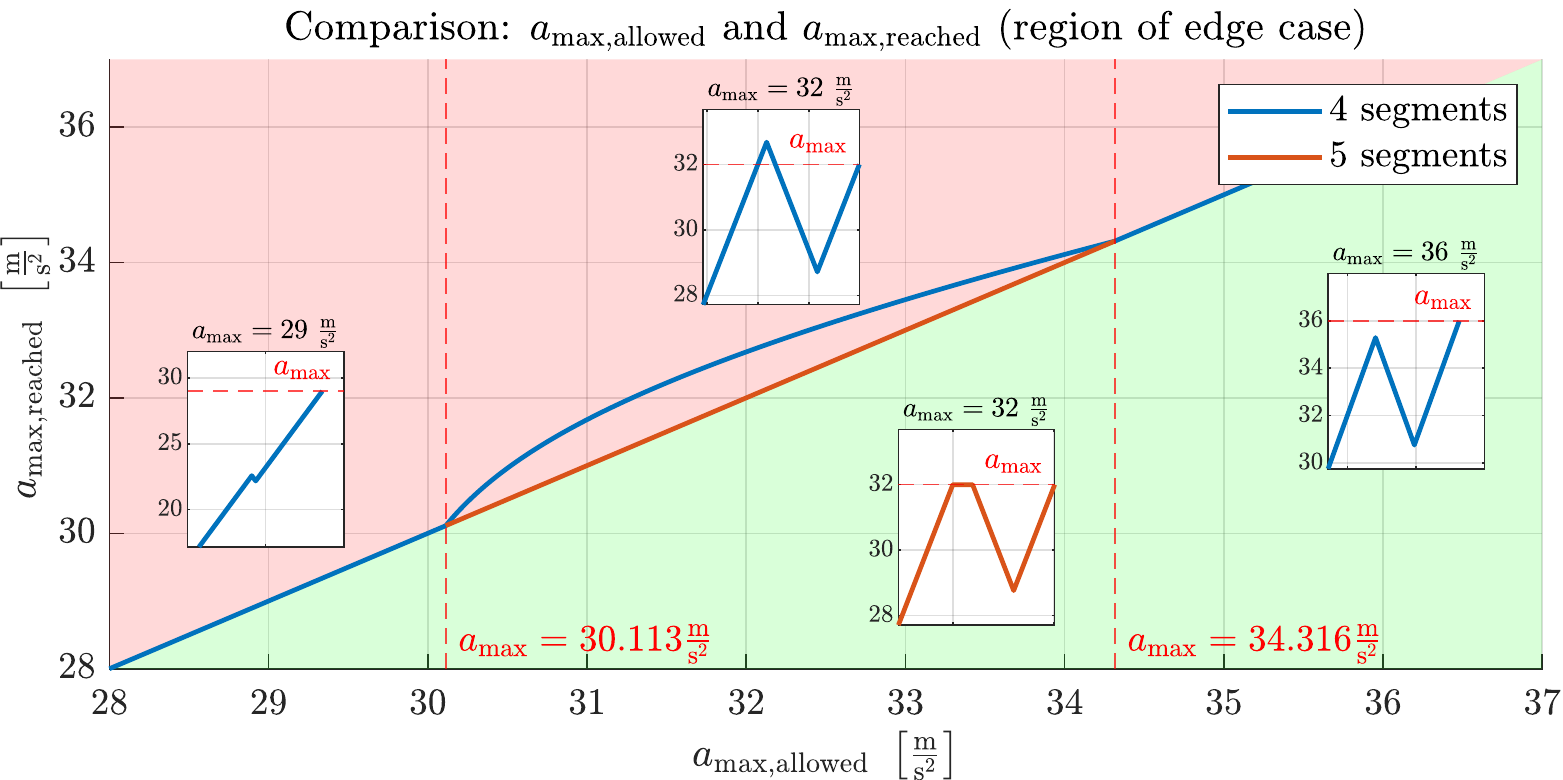 %
	\caption{Showing, the maximum acceleration on $t\in\left[0, t_\text{f}\right]$ and showing the interval, where the edge case can occur.}%
\label{fig:showing_the_edge_case_occuring}
\end{figure}

With the algorithm presented in this section, the jerk segments required to assemble the trajectory (as demonstrated in \cite{tau_ocpJ_assembly_part1}) can be calculated. The optimal control problem is formulated in a way, that $a_\text{max}$ is reached at the terminal time $t_\text{f}$ of the jerk segment. Since the acceleration is unconstrained on $t \in \left(0, t_\text{f}\right)$, $\ddot{z}\ttt$ can be higher than $\ddot{z}\tft$, as shown in \autoref{fig:showing_the_edge_case_occuring}.
This happens only on a small interval and, as \autoref{fig:graphical_representation_to_solve_ocpJ} and \autoref{fig:exemplary_solution_lambda1_and_5} show, it does not come into effect for $a_\text{max} = 2\cdot a_\text{lim}$ (required for the trajectory shown on the left in \autoref{fig:two_OcpJ_trajectories_for_demo_of_method}) for the parameters used in this publication. Introducing another $5^\text{th}$ step into the jerk segment (with this a section with $j\ttt = 0$) can help to remedy this issue as shown in \autoref{fig:showing_the_edge_case_occuring}. Implementation into the algorithm, further analysis when it appears and implementation details of the required calculation is scope for planned future work. Currently, violations caused by this are handled by the overlying algorithm of the $\ocpJ$ framework by reducing and optimizing (see Algorithm~2 and Algorithm~3 in \cite{tau_ocpJ_assembly_part1}).

\section{Comparison to other approaches}\label{sec:comp_to_others}
\begin{figure}[!ht]
	\centering
	\def\svgwidth{0.95\linewidth}
	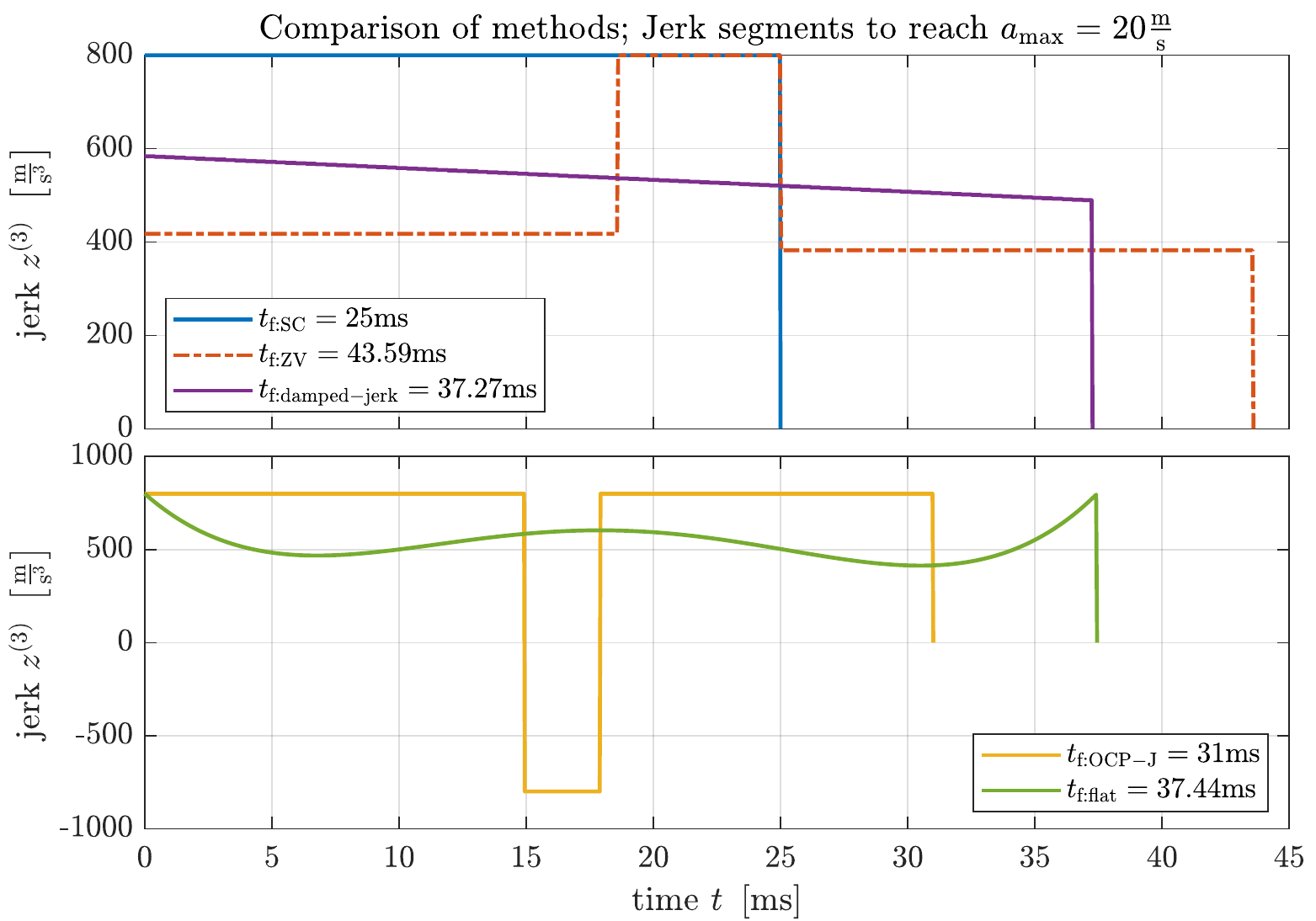 %
	\caption{Comparison of the different approaches to reach the same acceleration $\left(a_\text{max} = \SI{20}{\meter\per\second\squared}\right)$.}
	\label{fig:exemple_jerk_segments_alternative_approaches_accel_lower}
\end{figure}

\begin{figure}[!ht]
	\centering
	\def\svgwidth{0.95\linewidth}
	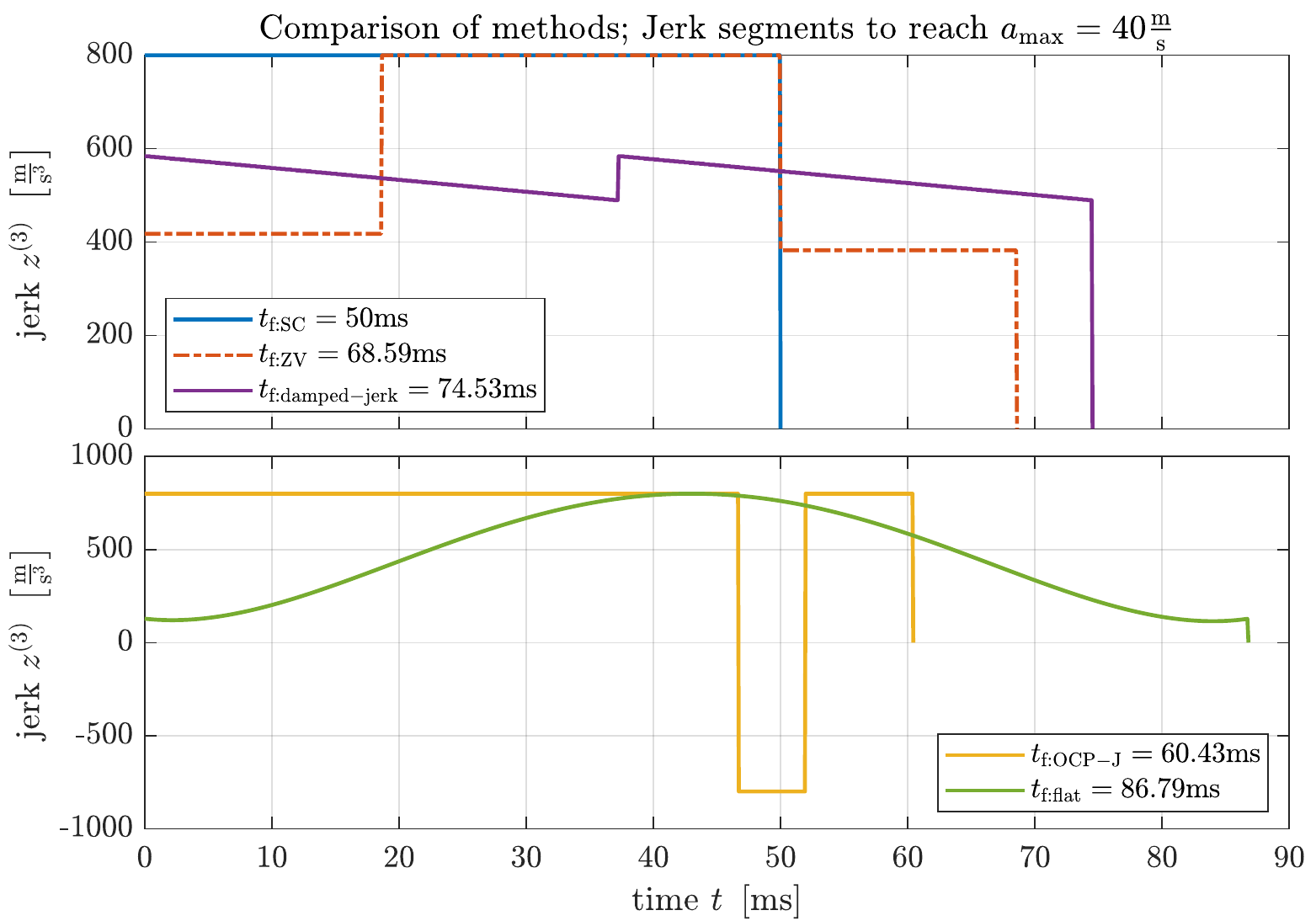 %
	\caption{Comparison of the different approaches to reach the same acceleration $\left(a_\text{max} = \SI{40}{\meter\per\second\squared}\right)$.}
	\label{fig:exemple_jerk_segments_alternative_approaches_accel_higher}
\end{figure}

\begin{figure}[!ht]
	\centering
	\def\svgwidth{0.95\linewidth}
	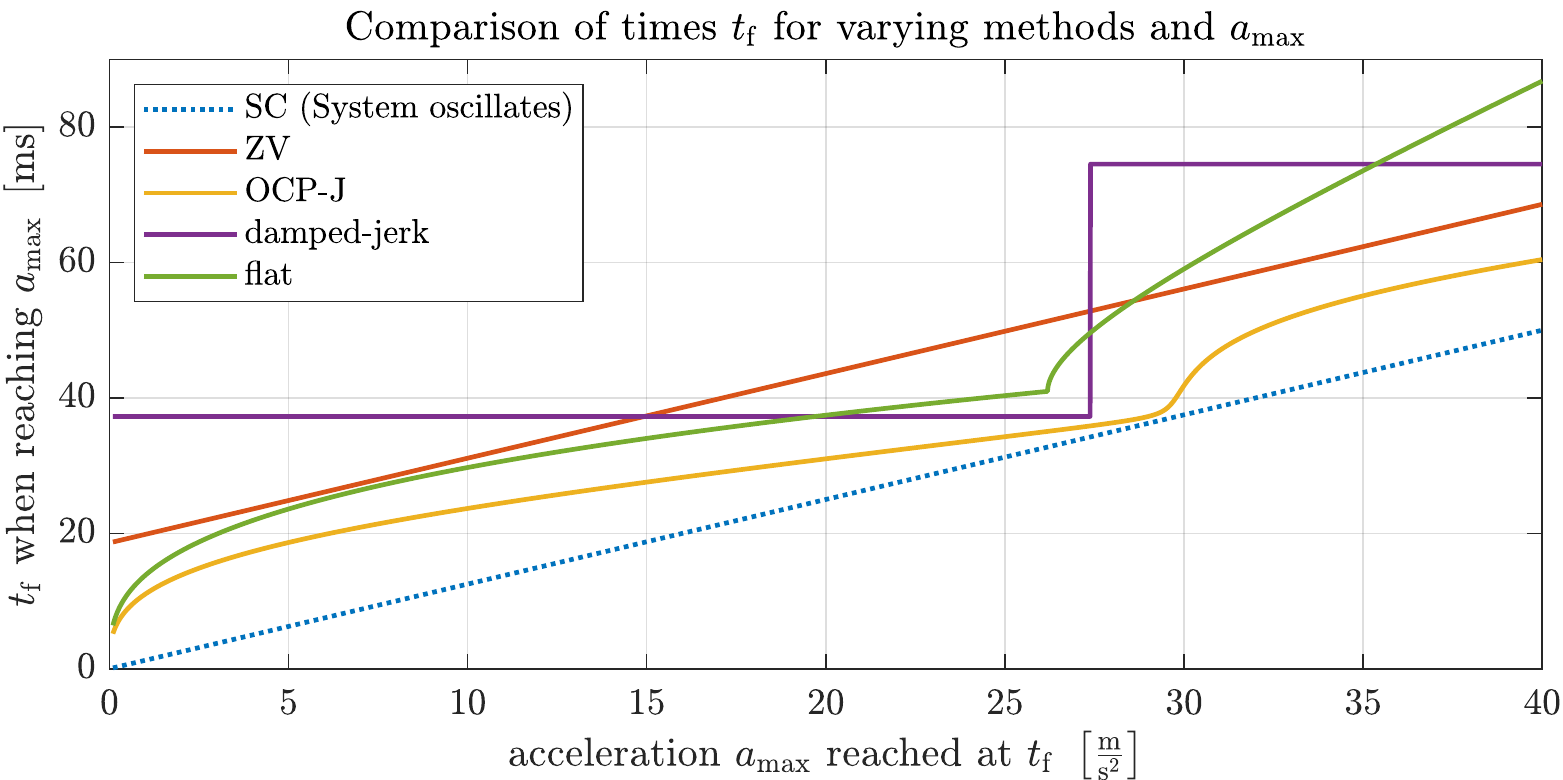 %
	\caption{Analysis of the different approaches over a bigger range of accelerations.}
	\label{fig:exemple_jerk_segments_alternative_approaches_accelSweep}
\end{figure}

Comparisons in this paper are made with the problem described in \autoref{SubSec:PMP_for_jerk_segments}. To summarize: The jerk segments must allow the slider acceleration to reach a certain value (labelled $a_\text{max}$), while ensuring that there is no oscillation in the base. The approach presented in this publication allows this to be done in a time-optimal manner. This section continues with a comparison with other established approaches, highlighting individual advantages and disadvantages. The first method that can be used and comes closest to the method presented is $\ZV$ shaping \cite{Singhose1990}, applied to a single jerk step as shown in \autoref{fig:exemple_jerk_segments_alternative_approaches_accel_lower}. The second method is a variation of a $\Fir$ shaper \cite{Biagiotti2015,Biagiotti2019,Biagiotti2020,Biagiotti2021,Yalamanchili2024}. For the comparison here, the variant presented in \cite{Bearee2014,Biagiotti2015} can be used, that allows to account for damping (which is only available for this kind of problem and has not been extended to a full trajectory like the $\Fir$ shapers presented in \cite{Biagiotti2019,Biagiotti2020,Biagiotti2021,Yalamanchili2024}). The final method capable of meeting the requirements is a flatness based approach \cite{Kronthaler2020IftoMM}. To provide a baseline, the jerk segments of the $\SCurve$ (utilising the maximum jerk, but ignoring the base oscillation requirements) are also given. All methods are shown for a single jerk segment in \autoref{fig:exemple_jerk_segments_alternative_approaches_accel_lower} for $a_\text{max} = \SI{20}{\meter\per\second\squared}$ and for  $a_\text{max} = \SI{40}{\meter\per\second\squared}$ in \autoref{fig:exemple_jerk_segments_alternative_approaches_accel_higher}.
Afterwards the analysis is expanded to more accelerations (sweep over a larger range of accelerations) and the resulting times $t_\text{f}$ are shown in \autoref{fig:exemple_jerk_segments_alternative_approaches_accelSweep}.
The advantage of the $\ocpJ$ approach over the other presented approaches lies in the reduced transition times. Since the method is time optimal for the problem presented, it offers the lowest transition time. The advantage of the other approaches presented however lies in the ease of implementation, since often only a few lines of code are required. The computation required by the $\ocpJ$ method presented in \autoref{sec:efficient_algorithm} is significantly more involved. The results are discussed further in \autoref{sec:discussion} and an analysis of how the times behave for the full trajectories can be found in \cite{tau_ocpJ_assembly_part1}.

\section{Measurements}\label{sec:measurements_lab_sys}
In order to validate the method presented, measurements were carried out on a laboratory system. In this section, the system is shown, the parameters and kinematic constraints are given and the measurement results are presented. The calculation of the full trajectories and the post-processing step to obtain the reference trajectory on the discrete controller intervals are mentioned.

\subsection{Laboratory system for measurements}
\begin{figure*}[!ht]
	\centering
	\def\svgwidth{0.95\linewidth}
	\footnotesize
	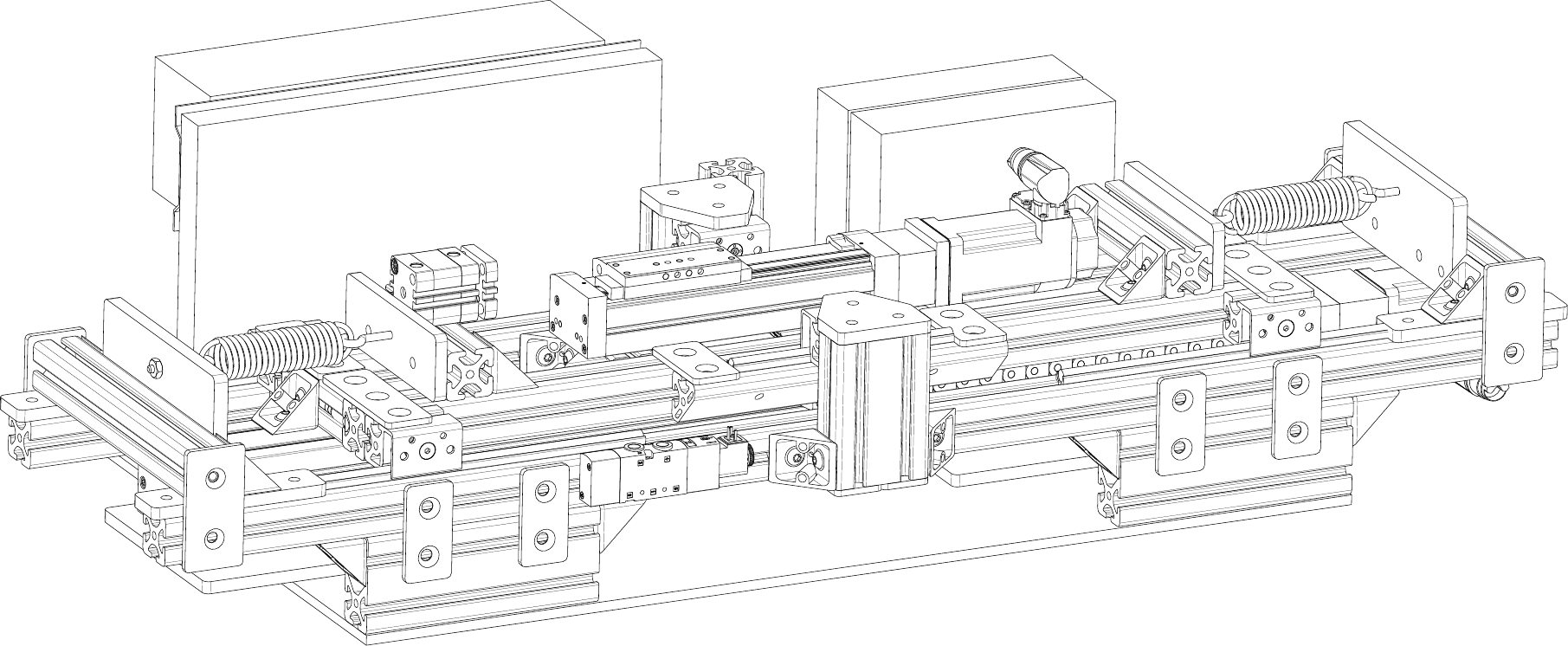
	\caption{CAD-picture of the laboratory system used for the measurements (picture also shown in \cite{Auer2024Case}).}
	\label{fig:CAD_of_laboratory_system_IACE}
\end{figure*}

A CAD-picture of the system is shown in \autoref{fig:CAD_of_laboratory_system_IACE}.
The main movement of the testrig is achieved by the mass located between the two springs. Its displacement corresponds to the variable $x\ttt$ in \autoref{sec:model_and_paramters}. Moreover, the position of the slider corresponds to $z\ttt$. The parameters and kinematic constraints of the laboratory system can be found in \autoref{tab:table_listing_machine_parameters_labSys}. 
\begin{table}[!ht]
	\captionsetup{width=\linewidth}
	\caption{Parameters, resulting eigenfrequencies and kinematic constraints of the laboratory system.}
	\vspace{-0.5em}
	\renewcommand{\arraystretch}{1.25}
	\centering
	\begin{tabular}{|l|l|l|}
		\hline
		$m_\text{s} = \SI{4.6546}{\kilo\gram}$ & $m_\text{b} = \SI{26.9057}{\kilo\gram}$ & $f_\text{0} = \SI{9.711}{\hertz}$ \\ \hline 
		$k = \SI{117499}{\newton\per\meter}$ & $d = \SI{50.4}{\kilo\gram\per\second}$ & $f_\text{d} = \SI{9.71}{\hertz}$ \\ \hline 
		$v_\text{max} = \SI{0.45}{\meter\per\second}$ & $a_\text{max} = \SI{6}{\meter\per\second\squared}$ & $j_\text{max} = \SI{200}{\meter\per\second\cubed}$ \\ \hline
	\end{tabular}
	\label{tab:table_listing_machine_parameters_labSys}
\end{table}
Using this system to validate the trajectory planning approach further serves to demonstrate, that the approach is not limited to the parameters given in \autoref{tab:table_listing_machine_parameters}.

\subsection{Calculation of trajectories and post-processing}
The calculation of the trajectories is done on the actual PLC. For this purpose, the calculation for the jerk segments presented in \autoref{sec:efficient_algorithm} and the calculations required for the trajectory assembly from \cite{tau_ocpJ_assembly_part1} have been implemented. All calculations refer to continuous time, so the times $t_n$ are not tied to a specific controller cycle. The system uses a position controller for the movement in order to track a given reference trajectory $t \mapsto z\ttt$ being computed according to \hyperref[App:slider_move_from_impulse]{Appendix~\ref*{App:slider_move_from_impulse}}. This allows to use the continuously calculated trajectory on the discrete controller interval ($t_\text{ctrl} = \SI{400}{\micro\second}$) and is a common approach \cite{ClaudioMelchiorri2008}. However, as a consequence the final time of a trajectory $t_\text{f,t}$ is always rounded up to a full controller cycle $t_\text{ctrl}$.

\subsection{Measurements of trajectories}
\begin{figure}[]
	\centering
	\def\svgwidth{0.95\linewidth}
	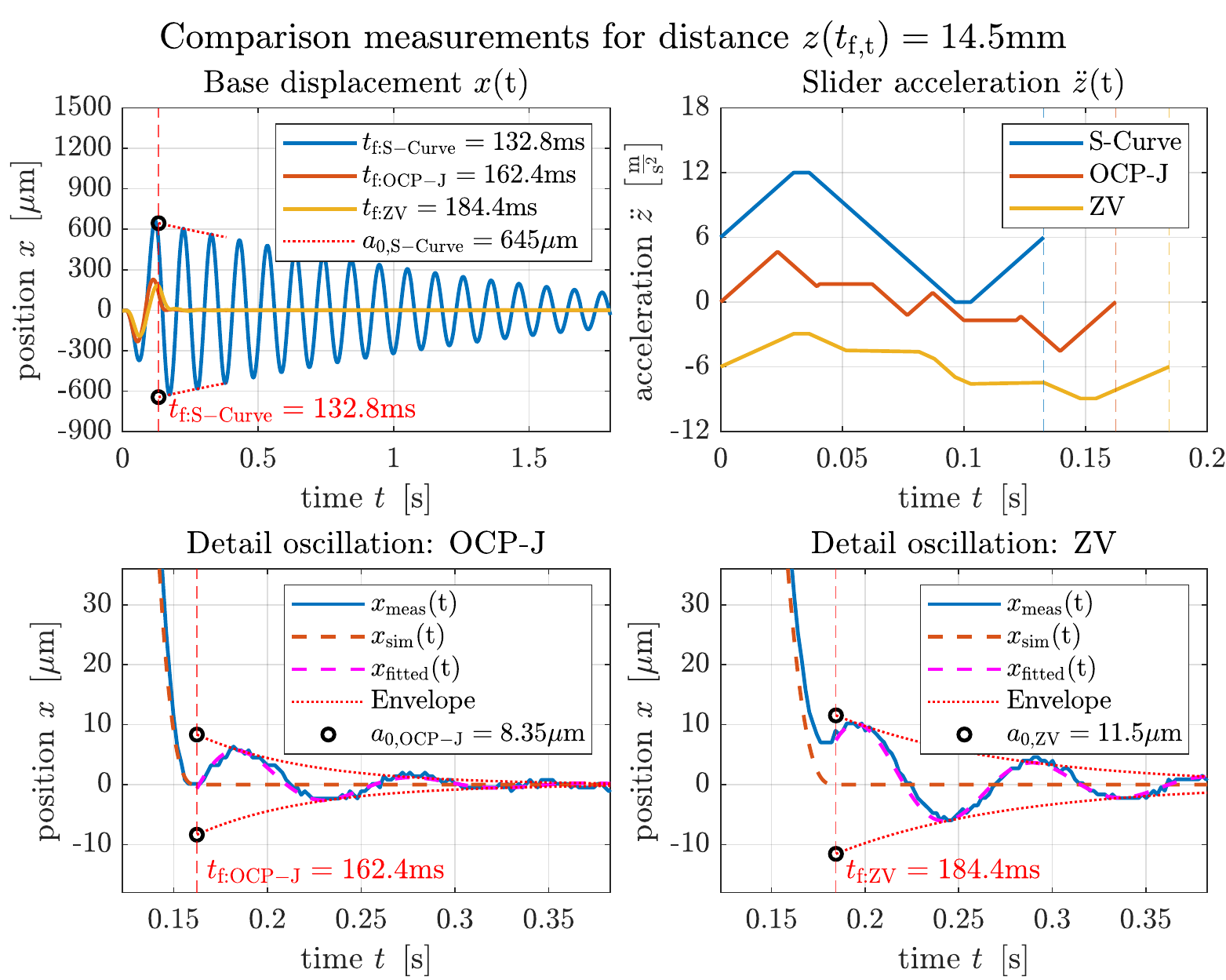%
	\caption{Response of the system to the trajectory when utilizing the presented approach for a transition distance of $\SI{14.5}{\milli\meter}$.}
	\label{fig:measurement_short_distance}
\end{figure}

\begin{figure}[]
	\centering
	\def\svgwidth{0.95\linewidth}
	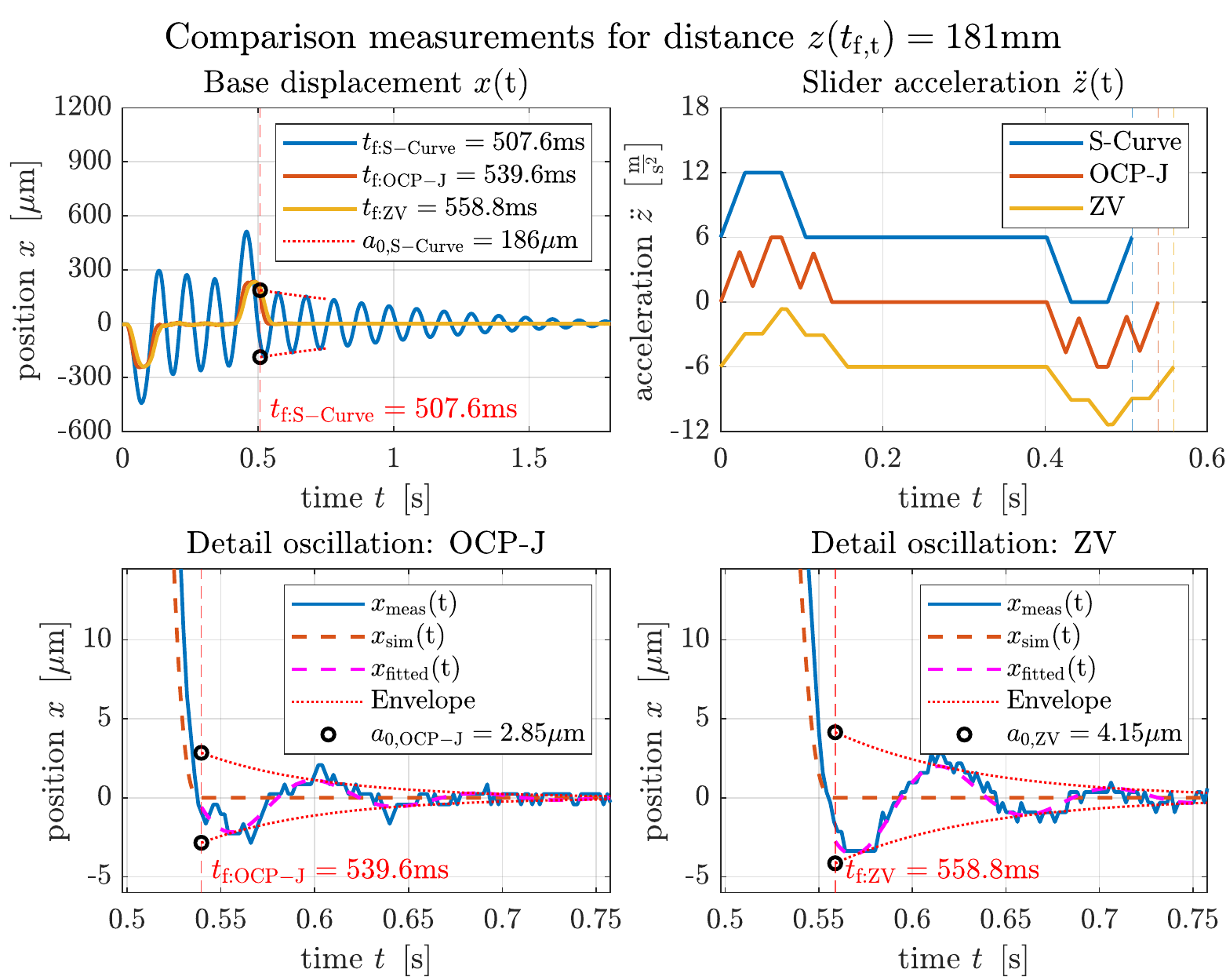%
	\caption{Response of the system to the trajectory when utilizing the presented approach for a transition distance of $\SI{181}{\milli\meter}$.}
	\label{fig:measurement_long_distance}
\end{figure}

\begin{figure}[]
	\centering
	\def\svgwidth{0.95\linewidth}
	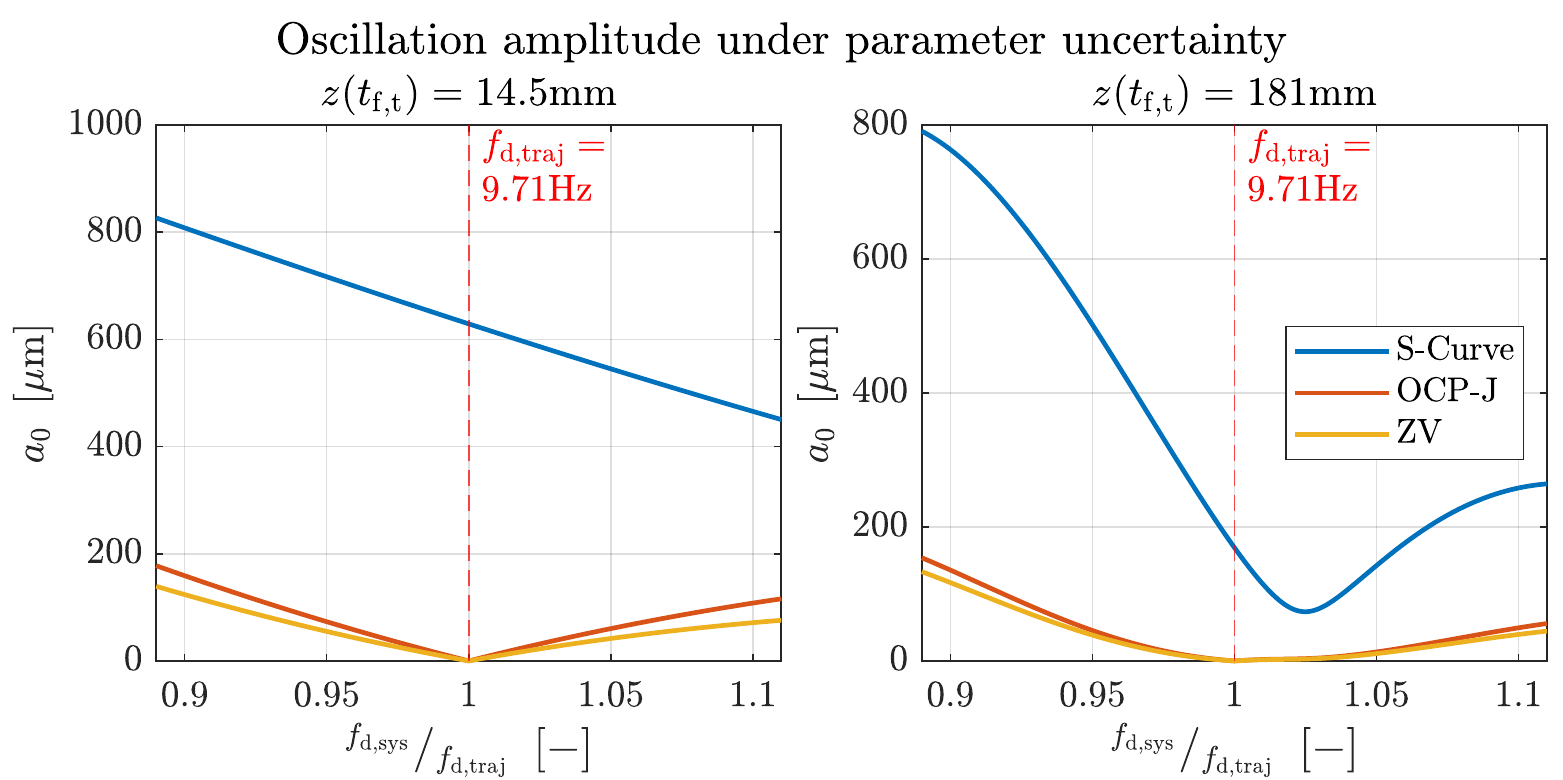%
	\caption{Amplitude of oscillation under larger parameter uncertainty for the distances shown in \autoref{fig:measurement_short_distance} and \autoref{fig:measurement_long_distance}.}
	\label{fig:parameter_uncertainty_shown_twoDist}
\end{figure}

A comparison with the $\SCurve$ is shown to demonstrate how significant the oscillation is without applying any trajectory planning approach. A measurement for a short transition of only $\SI{14.5}{\milli\meter}$ is shown in \autoref{fig:measurement_short_distance} and for a longer transition distance of $\SI{181}{\milli\meter}$ in \autoref{fig:measurement_long_distance}. As shown in \cite{Auer2023IFAC,Auer2024Case}, the amplitude of base oscillation (at $t_\text{f,t}$) depends on the overall distance. For the first distance, shown in \autoref{fig:measurement_short_distance}, the oscillation is quite substantial while for the one, shown in \autoref{fig:measurement_long_distance}, the amplitude is significantly smaller, as the measurements show. In addition, the trajectory corresponding to the shorter distance exhibits an overlap of the segments similar to that shown in \autoref{fig:showing_overlapping_1p5mm}. Since the kinematic constraints and system parameters are different, the exact distance for a similar overlap is also different (here it is $z\tftt = \SI{14.5}{\milli\meter}$ instead of $z\tftt = \SI{1.5}{\milli\meter}$ as in \autoref{fig:showing_overlapping_1p5mm}). The second trajectory corresponding to the transition distance of $\SI{181}{\milli\meter}$ is closer to the one shown in \autoref{fig:two_OcpJ_trajectories_for_demo_of_method} on the right (transition for a larger distance where no $a_\text{max} = a_\text{lim}$ is used directly).
The amplitude of oscillation $a_0$ as given in the plot is determined, by fitting an oscillation conforming to
\begin{equation}
	x_\text{fitted}\! \left(\tau\right) = \underbrace{a_0 \cdot e^{-\delta \, \tau}}_{\text{Envelope}} \, \sin \! \left( \omega_{\text{d}} \, \tau + \varphi_0 \right)
\end{equation}
to the measurement data $x_\text{meas}\ttt, t \ge t_\text{f,t}$. This is visualized in the lower to subplots, where the residual amplitudes for the different approaches are given. Theoretically, the oscillation should be zero after $t_\text{f,t}$, as the comparison with the simulation $x_\text{sim}\ttt$ shows. However, model and parameter uncertainties lead to the residual oscillation as shown in the plots. Considering that the amplitude obtained with $\SCurve$ is about $50$ times larger than the residual amplitude $a_0$ when using a trajectory planning approach, the measurements validate the approach. It is important to note that although the measured amplitude in the plots is smaller for the $\ocpJ$ trajectories here, the approach is not as robust for larger parameter uncertainties as the $\ZV$-shaper. This is apparent from the simulation results for both distances in \autoref{fig:parameter_uncertainty_shown_twoDist}.
A detailed study \wrt parameter uncertainty, comparisons with different trajectory planning methods, all validated by measurements, can be found in \cite{tau_ocpJ_assembly_part1}. %

\section{Discussion}\label{sec:discussion}
The trajectory planning method $\ocpJ$, as introduced in \autoref{sec:calc_full_trajectory} (visualized in \autoref{fig:two_OcpJ_trajectories_for_demo_of_method}) benefits from an algorithm that allows fast recalculation of individual jerk segments. This helps to avoid violations of kinematic constraints and ensures minimal transition times for the method, as shown in \autoref{fig:showing_overlapping_1p5mm}. It is advantageous to calculate the jerk segments in such a way that the acceleration is reached in minimum time in order to reduce the overall transition time. A comparison with other approaches from the literature is shown in \autoref{sec:comp_to_others} and the results are discussed here. The main advantage of the formulation given and solved in \autoref{sec:efficient_algorithm} is its inherent time optimality. Compared to the established $\ZV$-approach \cite{Singhose1990}, a significant gain in transition time can be achieved. Several methods \cite{Singhose1990,Bearee2014,Biagiotti2015,Biagiotti2019,Biagiotti2020,Biagiotti2021,Yalamanchili2024,Kronthaler2020IftoMM} can be adapted to the problem given in \autoref{sec:efficient_algorithm}, leading to the comparison shown in \autoref{fig:exemple_jerk_segments_alternative_approaches_accelSweep}. While the $\ocpJ$ method presented gives the fastest jerk segments, the computational effort is also the highest of the methods presented. The calculation for most of the other methods presented is simpler, while $\ocpJ$ requires a more extensive implementation. It is important to note, however, that the calculation time for $\ocpJ$ is completely predictable, as exactly the same steps are to be taken each time. While damping can be considered for a single jerk segment with \cite{Bearee2014}, the approach has not been extended to the full $\Fir$ methods as presented in \cite{Biagiotti2015,Biagiotti2019,Biagiotti2020,Biagiotti2021,Yalamanchili2024}. The fact that damping has been taken into account in the $\ocpJ$ approach allows it to be used on systems with damping. A detailed measurement study comparing the full trajectories (and not just the jerk segment) can be found in \cite{tau_ocpJ_assembly_part1}.

\bigskip
\subsection{Discussion of measurements}
Because a laboratory system with real springs instead of a steel structure (as you would find on a real machine) is used, the absolute amplitude measured on the laboratory system is naturally higher than the amplitude measured on a comparable pick-and-place machine. While both the $\ocpJ$ and $\ZV$-shaping approaches allow trajectories to be planned without oscillation for damped systems, some oscillation is unavoidable due to small parameter and model uncertainties. The fact that the amplitude of oscillation is lower for the $\ocpJ$ trajectories for the two measurements shown is just a coincidence. In general, as shown in \autoref{fig:parameter_uncertainty_shown_twoDist} and discussed in detail in \cite{tau_ocpJ_assembly_part1}, the sensitivity to parameter uncertainty is slightly higher for the $\ocpJ$ approach compared to $\ZV$-shaping. Further measurements and a detailed comparison with other trajectory planning approaches are provided in \cite{tau_ocpJ_assembly_part1}. Since this publication focuses on the calculation of the jerk segments, it was deemed sufficient to demonstrate the validity of the trajectories with the measurements shown in \autoref{fig:measurement_short_distance} \ref{fig:measurement_long_distance}.

\bigskip
\subsection{Applicability to other systems and parameter selection}
A concrete example of a system with system parameters and kinematic constraints was given in \autoref{sec:model_and_paramters}. However, it is possible to apply the approach to other oscillating systems, since the calculation only relies on the kinematic constraints, the oscillation frequency $\omega_{\text{d}}$ and damping $\delta$ of the system. These are basically the same parameters required to calculate a $\ZV$-shaped trajectory (which is still very relevant \cite{Kruk2023}). As far as the kinematic constraints are concerned, the velocity $v_\text{lim}$ and acceleration $a_\text{lim}$ result from the drives themselves. The kinematic constraint for the jerk $j_\text{lim}$ usually stems from prior experience on systems, since higher jerk generally leads to higher vibration (see \autoref{eq:x_to_xpp_sum_from_impulses} and \cite{Hayati2020,Dai2020}) and thus to increased wear \cite{Esau2022,Bilal2023}. If no such experience is available, a reasonable limit for the jerk can be set as follows
\begin{equation}
	j_\text{lim,init} = \frac{a_\text{max} \cdot \omega_{\text{d}}}{2 \, \pi} \, \text{.}
\end{equation}
Adjustments can be made from this limit depending on the response of the system (reduce jerk to improve slider tracking and thus reduce system vibration and wear, or increase jerk if there are no problems to reduce transition times). This should allow the approach to be used on a system even if there is no previous experience \wrt limiting the jerk $j_\text{lim}$. In addition to the system parameters, the maximum number of iterations of the line search method $n_\text{iter,max}$ (see \autoref{alg:Calc_jerk_segment}) can be adjusted. The maximum number of iterations leading to an improvement was given at the end of \autoref{SubSec:Efficient_algorithm_numerically}. Reducing the number of iterations reduces the computation time, but results in a small error. Since slider tracking errors and unavoidable small parameter uncertainties have the same effect, a reduction can be feasible.

\bigskip
\subsection{Limitations of the approach}
A potential limitation compared to other methods may be the implementation effort of the required methods. A measure of computational time has been given (see end of \autoref{SubSec:Efficient_algorithm_numerically}). As \cite{tau_ocpJ_assembly_part1} shows, the presented approach offers lower oscillations than $\Fir$-based methods \cite{Biagiotti2015,Biagiotti2019,Biagiotti2020,Biagiotti2021,Yalamanchili2024} and allows to consider the system damping directly. Compared to these methods, it also offers improved behaviour with respect to parameter uncertainty. In addition to the implementation effort, a limitation of the method presented here may be its sensitivity to parameter uncertainty, which is slightly worse than that of $\ZV$-shaping (see: \autoref{fig:parameter_uncertainty_shown_twoDist} and \cite{tau_ocpJ_assembly_part1}), but in general better than $\Fir$ \cite{tau_ocpJ_assembly_part1}. If the current method used in a system is $\ZV$ shaping, and system oscillations (due to unavoidable model and parameter uncertainties) are not considered an issue, then the $\ocpJ$ method presented here can be considered a valid replacement to reduce the overall transition time.

\section{Conclusion and outlook}\label{sec:summary_and_comparison}
A method has been introduced that allows for fast and predictable calculation of jerk segments. The formulation ensures that the time-optimal solution (for the jerk segments) is computed after a fixed number of steps, independent of system parameters and kinematic constraints. These jerk segments can be further assembled into a full trajectory, as presented in \cite{tau_ocpJ_assembly_part1}, to allow for near time optimal point-to-point transitions for the full trajectory $t_\text{f,t}$. The calculation does not rely on complex numerical optimisation procedures or algorithms, thus allowing implementation on a PLC. An extensive measurement and simulation study, focusing on the influence of parameter uncertainty and including further comparisons with other approaches, can be found in \cite{tau_ocpJ_assembly_part1} alongside case studies with different sets of parameters and kinematic constraints. Measurements from a laboratory system, as shown in \autoref{sec:measurements_lab_sys}, with trajectories calculated directly on the PLC are used to validate the approach and demonstrate its effectiveness. Another conceivable application of the jerk segments as motion primitives would be in a graph-based trajectory planning algorithm, such as the rapid exploration of random trees \cite{LaValle2009}, to plan trajectories in multi-dimensional configuration spaces. Possible extensions of the algorithm to consider multiple eigenfrequencies may pave the way for application to systems with multiple elastic modes. This is not as simple as simply replacing the system model \eqref{eq:opt:sys}, since the approach presented in \autoref{SubSec:Introduction_graphical_approach} is tailored to the particular model. Its extension to systems with more complex internal dynamics is therefore not straightforward and requires considerable research effort.

\appendices
\section{Slider movement for piecewise constant jerk $z^{\left(3\right)}$}\label{App:slider_move_from_impulse}
Trajectories of the form \eqref{eq:trajectory_jerk} with $n$ jumps in the jerk $z^{\left(3\right)}\ttt$ of amplitudes $a_1,\dots,a_n$ occurring at $t_1<\dots<t_n$ are considered.
The acceleration \eqref{eq:calc_zpp_from_imp}, velocity \eqref{eq:calc_zp_from_imp} and position \eqref{eq:calc_z_from_imp} can be calculated by integrating and taking the initial conditions values into account. This leads to
\begin{subequations}\label{eq:calc_z_to_zppp_from_snap}
	\begin{align}
				\ddot{z}\ttt &= \ddot{z}_0 + \sum\limits_{i=1}^{n} a_i \cdot \left(t - t_i\right) \, \text{,} \label{eq:calc_zpp_from_imp} \\
				\dot{z}\ttt &= \dot{z}_0 + \ddot{z}_0 \, t + \frac{1}{2} \, \sum\limits_{i=1}^{n} a_i \cdot \left(t - t_i\right)^2 \, \text{,} \label{eq:calc_zp_from_imp} \\
				z\ttt &= z_0 + \dot{z}_0 \, t + \ddot{z}_0 \, \frac{t^2}{2} + \frac{1}{6} \, \sum\limits_{i=1}^{n} a_i \cdot \left(t - t_i\right)^3 \label{eq:calc_z_from_imp}
			\end{align}
\end{subequations}
with $\ddot{z}_0$, $\dot{z}_0$ and $z_0$ being the initial values.

\section{Base movement for piecewise constant jerk $z^{\left(3\right)}$}\label{App:sys_answer_to_trajectories}
Generally, trajectories with jumps in the jerk $z^{\left(3\right)}\ttt$, as given in \hyperref[App:slider_move_from_impulse]{Appendix~\ref*{App:slider_move_from_impulse}}, are analysed. The calculations shown here all show the result for the base motion $x$, $\dot{x}$, $\ddot{x}$ and $x^{\left(3\right)}$ to an impulse in the snap $z^{\left(4\right)}$ with length $a_i$.
The damping $\delta$ of the system and the damped angular frequency $\omega_\text{d}$ are calculated with
\begin{align}\label{eq:calculate_delta_w0_wd_from_parameters}
	\delta &= \frac{d}{2\, m_\text{g}} \, \text{,} & \omega_0^2 &= \frac{k}{m_\text{g}} \, \text{,} & \omega_\text{d} &= \sqrt{\omega_0^2 - \delta^2} \, \text{.}
\end{align}
Assuming, that $x_0$ and all its derivatives vanish at $t=0$, taking multiple impulses in $z^{\left(4\right)}$ into account (see \eqref{eq:trajectory_jerk}) and using $\tau_i = t - t_i$, gives the symbolic solution for the base motion
\begin{subequations}\label{eq:x_to_xpp_sum_from_impulses}
	\begin{align}
		& \ddot{x}\ttt = \sum\limits_{i=1}^{n} -a_i \frac{m_\text{s}}{m_\text{g} \, \omega_\text{d}} \, e^{-\delta \, \tau_i} \, \sin \! \left(\omega_\text{d} \, \tau_i\right) \cdot H \! \left(\tau_i\right) \, \text{,} \label{eq:xpp_sum_from_impulses} \\
		& \begin{aligned}
			\dot{x}\ttt = \sum\limits_{i=1}^{n} a_i & \frac{m_\text{s}}{m_\text{g} \, \omega_\text{d} \, \omega_\text{0}^2} \Big[e^{-\delta \, \tau_i}\cdot\big(\omega_\text{d} \, \cos \! \left(\omega_\text{d} \, \tau_i\right) + \\
			& \delta \, \sin \! \left(\omega_\text{d} \, \tau_i\right) \big) - \omega_\text{d}  \Big] \cdot H \! \left(\tau_i\right) \, \text{,} \label{eq:xp_sum_from_impulses}
		\end{aligned} \\
		& \begin{aligned}
			x\ttt =  & \sum\limits_{i=1}^{n} a_i  \frac{m_\text{s}}{m_\text{g} \, \omega_\text{d} \, \omega_\text{0}^4}\! \Big[ e^{-\delta \, \tau_i}\cdot\Big( \left(\omega_\text{d}^2 - \delta^2\right) \, \sin \! \left(\omega_\text{d} \, \tau_i\right) \\
			& - 2 \, \omega_\text{d} \, \delta \, \cos \! \left(\omega_\text{d} \, \tau_i\right) \Big) - \omega_\text{d} \, 	\omega_\text{0}^2 \, \tau_i + 2 \, \delta \, \omega_\text{d} \Big] H \! \left(\tau_i\right) \, \text{.}  \label{eq:x_sum_from_impulses}
		\end{aligned}
	\end{align}
\end{subequations}
The base velocity \eqref{eq:xp_sum_from_impulses} and base position \eqref{eq:x_sum_from_impulses} are calculated via integration from the acceleration \eqref{eq:xpp_sum_from_impulses}. Deriving \eqref{eq:xpp_sum_from_impulses} with respect to the time gives the jerk of the base
\begin{equation}
	\begin{aligned}
		x^{\left(3\right)}\ttt = \sum\limits_{i=1}^{n} -a_i & \frac{m_\text{s}}{m_\text{g} \, \omega_\text{d}} \Big[e^{-\delta \, \tau_i}\cdot\big(\omega_\text{d} \, \cos \! \left(\omega_\text{d} \, \tau_i\right) + \\
		& \delta \, \sin \! \left(\omega_\text{d} \, \tau_i\right) \big) \Big] \cdot H \! \left(\tau_i\right) \, \text{.} \label{eq:xppp_sum_from_impulses}
	\end{aligned}
\end{equation}

\end{document}